\documentclass[prx,aps,showpacs,twocolumn,notitlepage]{revtex4-2}

\def\be{\begin{equation}}
\def\ee{\end{equation}}
\usepackage{graphicx}
\usepackage{bm}
\usepackage[latin9]{inputenc}
\usepackage{amsmath}
\usepackage{amssymb}
\usepackage{float}
\usepackage{lineno}
\usepackage[bookmarks=true,
   colorlinks=true,
   linkcolor=blue,
   urlcolor=blue,
   citecolor=blue,
   bookmarks=true,
   hyperindex=true
]{hyperref}
\usepackage{natbib}

\begin{document}
% \linenumbers

\title{Persistent breather and dynamical symmetry in a unitary Fermi gas}

\author{Dali Sun$^{1}$}
\thanks{These authors contributed equally to this work}
\author{Jing Min$^{1,4}$}
\thanks{These authors contributed equally to this work}
\author{Xiangchuan Yan$^{1}$}
\thanks{These authors contributed equally to this work}

\author{Lu Wang$^{1,4}$, Xin Xie$^{1,4}$, Xizhi Wu$^{1,4}$}

\author{Jeff Maki$^{5}$, Shizhong Zhang$^{2}$}

\author{Shi-Guo Peng$^{1}$}
\thanks{Corresponding author: pengshiguo@wipm.ac.cn}

\author{Mingsheng Zhan$^{1,3}$}

\author{Kaijun Jiang$^{1,3}$}
\thanks{Corresponding author: kjjiang@wipm.ac.cn}

\affiliation{$^{1}$State Key Laboratory of Magnetic Resonance and Atomic and Molecular Physics, Innovation Academy for Precision Measurement Science and Technology, Chinese Academy of Sciences, Wuhan 430071, China}
\affiliation{$^{2}$Department of Physics and Hong Kong Institute of Quantum Science and Technology, The University of Hong Kong, Hong Kong, China}
\affiliation{$^{3}$Wuhan Institute of Quantum Technology, Wuhan 430206, China}
\affiliation{$^{4}$University of Chinese Academy of Sciences, Beijing 100049, China}
\affiliation{$^{5}$ Pitaevskii BEC Center, CNR-INO and Dipartimento di Fisica, Universit\`a di Trento, 38123 Trento, Italy}

\date{\today}

\begin{abstract}
SO(2,1) dynamical symmetry makes a remarkable prediction that the breathing oscillation of a scale invariant quantum gas in an isotropic harmonic trap is isentropic and can persist indefinitely. In 2D, this symmetry is broken due to quantum anomaly in the strongly interacting range, and consequently the lifetime of the breathing mode becomes finite. 
%Although the persistent breather has been observed in noninteracting thermal gases, 
The persistent breather in a strongly interacting system has so far not been realized. Here we experimentally achieve the long-lived breathing mode in a 3D unitary Fermi gas, which is protected by the SO(2,1) symmetry. The nearly perfect SO(2,1) symmetry is realized by loading the ultracold Fermi gas in an isotropic trap and tuning the interatomic interaction to resonance. The breathing mode oscillates at twice the trapping frequency even for large excitation amplitudes. %, going beyond the linear regime. 
The ratio of damping rate to oscillation frequency is as small as 0.002, providing an interacting persistent breather. The oscillation frequency and damping rate keep nearly constant for different atomic densities and temperatures, demonstrating the robustness of the SO(2,1) symmetry in 3D. The factors that lead to the residual damping have also been clarified. This work opens the way to study many-body non-equilibrium dynamics related to the dynamical symmetry.
\end{abstract}

\maketitle

\section{Introduction}
Understanding nonequilibrium dynamics in quantum many-body systems is one of the most challenging problems in modern physics~\cite{Michael2018Nature, Langen2015}. A particularly interesting question is the search for long-lived far-from-equilibrium steady states and to understand whether such states have universal properties \cite{Schmiedmayer2018Nature, Oberthaler2018Nature, Hadzibabic2018Nature}.
In this regard, one famous example is the Boltzmann breather: non-interacting particles placed in an isotropic harmonic trap potential can undergo undamped oscillatory motion. Such motion was predicted using kinetic theory, but was only recently observed in thermal gases~\cite{Lobser2015}. However, observation of persistent breather in strongly interacting quantum systems is very challenging and has so far not been realized.
%well above the Bose-Einstein condensate transition temperature .

\begin{figure*} [htbp]
\includegraphics[width=17cm]{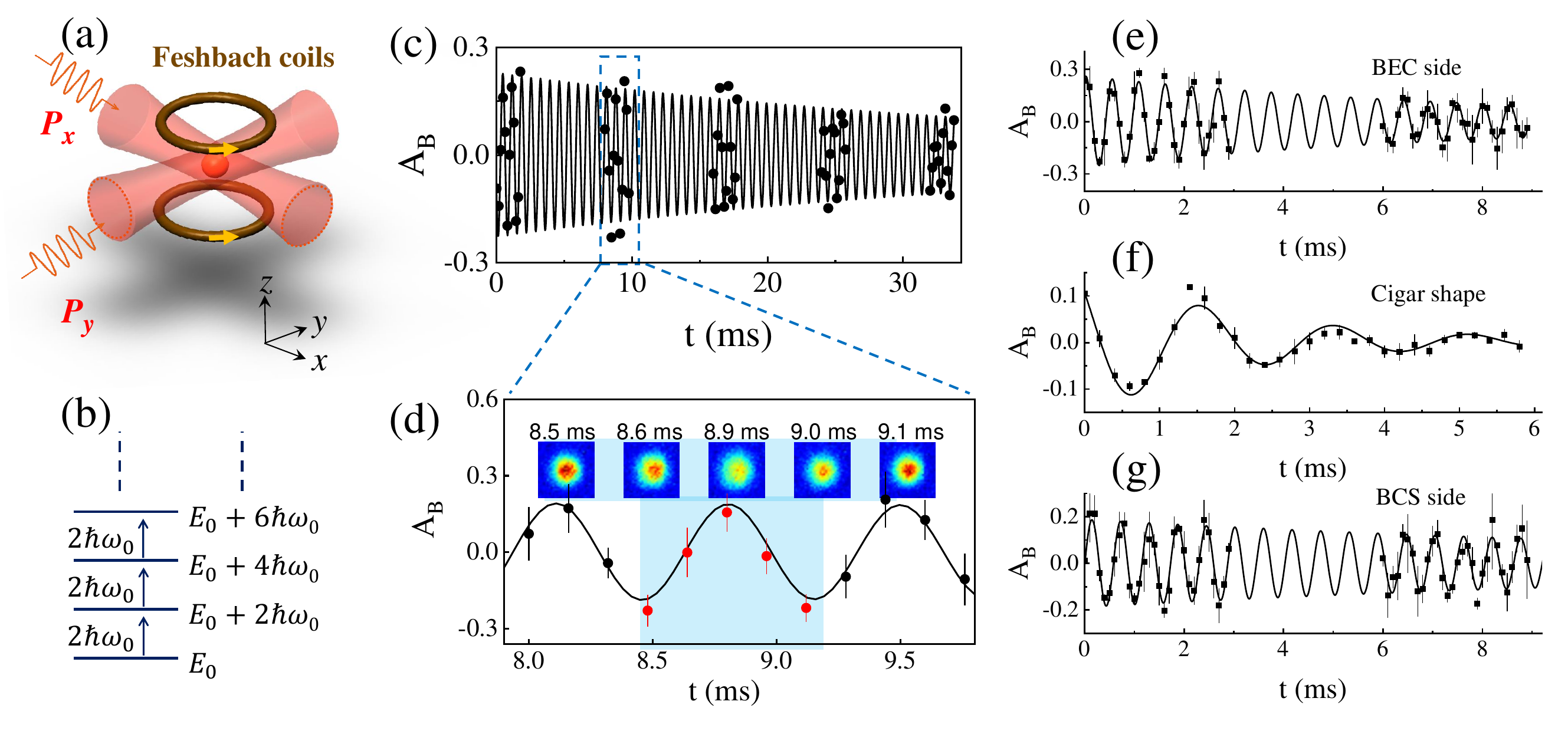}
\caption{Long-lived breathing mode of a unitary Fermi gas in an isotropic trap. (a) The isotropic optical dipole trap is composed of two elliptic laser beams ($P_{x}$ and $P_{y}$). A pair of Feshbach coils produces a homogeneous magnetic field to tune the interaction to the unitary limit.
%The breathing mode is excited by modulating the two beam powers according to the mode symmetry.
(b) Energy spectrum of the system. The ground state is $|\psi_{0}\rangle$ with the energy $E_{0}$. The raising operator $L_{+}$ produces an excitation with an energy $2\hbar\omega_{0}$. Repeated action of $L_{+}$ generates a series of evenly spaced states known as a conformal tower. (c) Breathing oscillation as a function of the holding time $t$. Enlarged view of the second data set is shown in (d), where the inset shows typical atomic images (166 $\times$ 166 $\mu m^{2}$) from 8.5 to 9.1 ms, corresponding to the red dots. The oscillation frequency is $\omega_{B}=2\pi\times1439(1)\text{ Hz}\approx 2\omega_{0}$ and the damping rate is $\Gamma_{B}/\omega_{B} \approx0.002$.
(e), (f) and (g) show the breathing modes with the SO(2,1) symmetry broken: on the BEC side ($1/k_{F}a\approx0.83$) in an isotropic trap (e), $\omega_{B} \approx2.11\omega_{0}$ and $\Gamma_{B}/\omega_{B} \approx 0.01$; at the unitary regime ($1/k_{F}a\rightarrow 0$) in a cigar-shaped trap (f), $\omega_{B} \approx 1.83\omega_{r}$ with the radial trapping frequency $\omega_{r}$ and $\Gamma_{B}/\omega_{B} \approx 0.13$; on the BCS side ($1/k_{F}a\approx -0.34$) in an isotropic trap (g), $\omega_{B} \approx1.96\omega_{0}$ and $\Gamma_{B}/\omega_{B} \approx 0.006$.
%(e) for the breathing oscillation on the BEC side [$1/(k_{F}a)\approx0.83$] in an isotropic trap. The oscillation frequency is $\omega_{B} = 2\pi\times1876\ \textrm{Hz} \approx2.11\omega_{0}$, the damping rate is $\Gamma_{B}= 117 \ s^{-1}$, and the normalized damping rate is $\Gamma_{B}/\omega_{B} \approx 0.01$; (e) for the radial breathing mode at unitarity [$1/(k_{F}a)\rightarrow 0$] in a cigar-shaped trap. The aspect ratio of the atomic axial radius to the radial one is $10:1$. The oscillation frequency is $\omega_{B} = 2\pi\times557\ \textrm{Hz} \approx \sqrt{10/3}\omega_{r}$ where $\omega_{r}$ is the radial trapping frequency, the damping rate is $\Gamma_{B}= 460 \ s^{-1}$, and the damping rate is $\Gamma_{B}/\omega_{B} \approx 0.13$; (g) for the breathing oscillation on the BCS side [$1/(k_{F}a)\approx -0.34$] in an isotropic trap. The oscillation frequency is $\omega_{B} = 2\pi\times1738\ \textrm{Hz} \approx1.96\omega_{0}$, the damping rate is $\Gamma_{B}= 64 \ s^{-1}$, and the normalized damping rate is $\Gamma_{B}/\omega_{B} \approx 0.006$.
The error bar is the standard deviation of three measurements. The oscillation data are fitted with a damped sinusoidal function indicated by the solid curve.}
\label{fig:Fig1}
\end{figure*}

SO(2,1) dynamical symmetry makes a remarkable prediction that the breathing oscillation is completely isentropic and can persist indefinitely \cite{Pitaevskii1997, Werner2006, Randeria2012Lecture, Ho2004, Son2007}. Mathematically, if scale invariant quantum gases are confined in an isotropic harmonic trap, the Hamitonian $H$ and raising/lowing operators $L_{\pm}$ form the SO(2,1) Lie algebra (see Sec. \ref{Sec:Appendix C}). Then the operators $L_{\pm}$ generate a set of equally spaced eigenstates known as conformal tower (see Fig. \ref{fig:Fig1}(b)): repeated action of $L_{+}$ on the ground state %with energy $E_0$ 
will generate a ladder of eigenstates with an equal energy separation $2\hbar \omega_{0}$, where $\omega_{0}$ is the trapping frequency. One important feature of SO(2,1) symmetry is that the breathing mode undergoes an undamped periodic oscillation with the frequency of $2\omega_0$ \cite{RadiusOscillation}. 
%the SO(2,1) symmetry leads to a set of conformal tower states; within each tower, the eigenstates are equally spced with spacing given by $2\hbar \omega_{0}$, where $\omega_{0}$ is the trapping frequency (see Fig. \ref{fig:Fig1}(b)). As a consequence, the mean-square radius of the gas, i.e. $\langle r^2 \rangle(t)$, undergoes undamped periodic oscillation with frequency $2\omega_0$, irrespective of the oscillation amplitude, temperature and density of the system~\cite{RadiusOscillation}.

In 2D quantum gases with the scale-invariant short-range interaction $g\delta^{2}(r)$, long-lived breathing motion was predicted from the SO(2,1) symmetry in the classic field obeying the nonlinear Gross-Pitaevskii (GP) equation \cite{Pitaevskii1997} and has been experimentally tested \cite{Vogt2012, Dalibard2002PRL, Saint2019}. However, such symmetry in 2D systems is broken due to quantum anomaly raised by the quantization of the short-range interaction \cite{Olshanii2010, Hofmann2012, Peppler2018, Holten2018, Murthy2019}. Especially in the strongly interacting regime, the quantum anomaly becomes non-negligible: it induces the up shift of the breathing-mode frequency away from the scale invariant one and increases the damping rate by about one order of magnitude \cite{Peppler2018, Holten2018}. Also with the increase of atomic number and temperature, the system evolves the dimensional crossover from 2D into 3D, and then suffers a rapid damping process \cite{Peppler2018, Holten2018}.

In a 3D quantum system, the asymptotic form of wave function with contact interaction is introduced as the Bethe-Peierls boundary condition, i.e., $\Psi(r\rightarrow 0)\sim (1/r-1/a)$, where $r$ is the distance between two atoms and $a$ is the scattering length. The interaction is scale invariant if the system is tuned to the unitary limit with resonant interaction $a=\pm \infty$ \cite{Werner2006, Ho2004, Son2007, Deng2016}. When the system is confined in an isotropic harmonic trap, the scale invariance is replaced by the SO(2,1) symmetry. In 3D, no quantum anomaly occurs at the resonant interaction and one realises a robust SO(2,1) symmetry~\cite{Werner2006, Randeria2012Lecture, Ho2004, Son2007}.  

%When a 3D quantum system is tuned to the unitary limit, the interaction is scale invariant \cite{Werner2006, Ho2004, Son2007, Deng2016}.
%Physically, this can be readily seen by noting that the operator for the mean-square radius is a superposition of $L_{\pm}$ operators, and as such only couple states within a given conformal tower.
%The mean-square radius of the gas satisfies the following equation of motion:
%\begin{equation}
%    \frac{d^2}{dt^2}\langle r^2 \rangle(t) +4 \omega_{0}^2 \langle r^2 \rangle(t) = \frac{4E}{m}
%    \label{eq:moi_eom}
%    \end{equation}
% where $E$ is the energy per atom and $m$ is the atomic mass.

In this work, we provide concrete evidence for the existence of a long-lived breather and of SO(2,1) symmetry in an isotropically trapped unitary Fermi gas. In particular, we excite and observe the long-lived breathing mode of a unitary Fermi gas. The main results are as follows: (i) This mode oscillates at twice the trapping frequency $\omega_{B} \approx 2\omega_{0}$ and with an extremely small damping rate $\Gamma_{B} \approx 0.002\omega_{B}$,  providing an interacting persistent breather (Sec. \ref{Sec:II});
 %obviously different from the case when the SO(2,1) symmetry is broken. 
(ii) The breathing mode oscillates at twice the trapping frequency even for large excitation amplitudes, going beyond the prediction of linear response theory  \cite{Castin1996, Stringari1996}, but consistent with the predictions of SO(2,1) symmetry \cite{Pitaevskii1997, Werner2006}, and is analogous of the Boltzmann breather in an interacting quantum system \cite{Maki2020} (Sec. \ref{Sec:III}); (iii) The oscillation frequency and damping rate keep nearly constant for different atomic densities and temperatures, demonstrating the robustness of the symmetry in 3D (Sec. \ref{Sec:IV});
%This unusual property of the undamped oscillation in a 3D quantum system has been theoretically predicted \cite{Werner2006, Randeria2012Lecture}.
(iv) We further analyze the factors which affect the residual damping rate, such as asphericity, anharmonicity and bulk viscosity, and find that the damping has reached the technical limit as characterized by the dipole mode. The asphericity is specially increased to study its effect on the damping (Sec. \ref{Sec:V}).

\section{Long-lived Breathing mode} \label{Sec:II}
We prepare a unitary $^{6}$Li Fermi gas in an isotropic trap as in our previous works \cite{Jiang2024PRL, Yan2021, Yan2022}. Two elliptic 1064-nm optical beams form the isotropic trap (see Fig. \ref{fig:Fig1}(a)). A pair of Feshbach coils produces a homogeneous magnetic field of 832 G to tune the Fermi gas with two balanced spin states $|F=1/2,m_{F}=\pm1/2\rangle$ into the unitary limit. The power of the isotropic optical trap is increased to $P_{0}$ for initially loading the cold atoms ($P_{x}=P_{y}=P_{0}$), then slowly decreased to $P_{1}$ for performing the evaporative cooling. Finally we adiabatically increase the power to $P_{2}$ for reducing the anharmonicity of the trap before exciting the breathing mode (see time sequence in Fig. \ref{fig:Sfig1}). The atomic temperature and density are controlled by optical powers $P_{1}$ and $P_{2}$, respectively. The atom number is $N=3\times10^{4}$ for typical experiment,
%the reduced temperature $T/T_{F}$ with the Fermi temperature $T_{F}$ varies from 0.29 to 0.44, the trapping frequency $\omega_{0}$ is  in the  range $2\pi\times (501$ $\sim$ 1121) Hz [i.e., the central atomic density $n_{0}$ in the range $(6\times10^{12} \sim 2\times10^{13}) \ \textrm{cm}^{-3}$],
and the asphericity $\delta\equiv \left(\omega_{\text{max}}-\omega_{\text{min}}\right)/\bar{\omega}_{0}$ is about $4.9\%$, where $\omega_{\text{max}}$, $\omega_{\text{min}}$, $\bar{\omega}_{0}$ are the maximum, minimum and geometric mean frequencies along three axes, respectively. To improve the stability of the system, we further reduce the uncertainty of the optical power to about 0.8$\%$ by using a PID feedback controller, and keep the surrounding temperature fluctuation less than 0.5 K. During the experiments, the fluctuations of the trap frequency and anisotropy are both smaller than 1.0$\%$.

%$\delta=\left(\omega_{\text{max}}-\omega_{\text{min}}\right)/\bar{\omega}_{0}\approx4.9\%$, where $\omega_{\text{max}}$, $\omega_{\text{min}}$, $\bar{\omega}_{0}$ are the maximum, minimum and geometric mean frequencies along three axes, respectively.

To excite the breathing mode, we sinusoidally modulate the optical field at frequency $2\omega_{0}$ for four periods, making the atomic cloud sizes oscillate in phase along three axes. After the modulation, we let the system evolve for a hold time $t$ and then image the cloud simultaneously along the $y$ and $z$ directions after a time of flight of 1 ms. The breathing mode oscillation is characterized by
\begin{equation} \label{eq:monopole}
 A_{B}\left(t\right) \equiv \frac{\sum_{i}\left\langle r_{i}^{2}\right\rangle(t)}{\left<\sum_{i}\left\langle r_{i}^{2}\right\rangle (t)\right>} -1, \ \ \left(i=x, y, z\right),
\end{equation}
where $\left\langle r_{i}^{2}\right\rangle (t)$ is the mean-square cloud radius in the $i$th direction and $\left<\sum_{i}\left\langle r_{i}^{2}\right\rangle (t)\right>$ is the average value of each data set (see Sec. \ref{Sec:Appendix A}). As shown in Fig. \ref{fig:Fig1}(c), the oscillation is fitted with a damped sinusoidal function $A_{B}=\delta A_{B}\sin\left(\omega_{B} t+\phi\right)\exp(-\Gamma_{B} t)$, where Fig. \ref{fig:Fig1}(d) shows the enlarged view of the second data set. The temperature is $T/T_{F}=0.29(1)$. Remarkably, with a large trapping frequency $\omega_{0}=2\pi\times716$ Hz, the oscillation can persist for tens of milliseconds. The oscillation frequency is $\omega_{B}=2\pi\times1439(1)\text{ Hz}\approx2.01\omega_{0}$, and the damping rate is $\Gamma_{B}=18(4)$ $\text{s}^{-1}$. The normalized damping rate to the oscillation frequency, i.e., $\Gamma_{B}/\omega_{B}$, is as small as 0.002, which is improved by a factor of two compared with the previous work \cite{Jiang2024PRL}. We note that the breathing mode of a 2D Fermi gas with the SO(2,1) symmetry has been measured at the University of Cambridge, and they obtained a normalized damping rate of 0.025 \cite{Vogt2012}.
%Similar experiments for the breathing mode have been conducted at Innsbruck \cite{altmeyer2007dynamics, altmeyer2007precision, riedl2008collective} and Duke \cite{kinast2005damping} for elongated superfluid Fermi gases. The obtained values for $\Gamma_{B}/\omega_{B}$ at the lowest recorded temperatures are about 0.006 and 0.008, respectively. We note also that the radial breathing mode of a 2D Fermi gas has been measured at the University of Cambridge, and they obtained a larger value of 0.025 \cite{Vogt2012}.
To the best of our knowledge, the normalized damping rate obtained in this work is the smallest for the strongly interacting Fermi gas \cite{Othergroup}.

For comparison, we also measure the breathing modes with the SO(2,1) symmetry broken (see Fig. \ref{fig:Fig1}(e), (f) and (g)): (1) on the BEC side ($1/k_{F}a\approx0.83$, $k_{F}$ is the Fermi wave vector and $a$ is the s-wave scattering length) in an isotropic trap, the oscillation frequency is $\omega_{B} \approx2.11\omega_{0}$ and the damping rate is $\Gamma_{B}/\omega_{B} \approx 0.01$; (2) at unitarity ($1/k_{F}a\rightarrow 0$) in a cigar-shaped trap whose aspect ratio is about 10, the oscillation frequency is $\omega_{B} \approx 1.83\omega_{r}$ ($\omega_{r}$ is the radial trapping frequency) and the damping rate is $\Gamma_{B}/\omega_{B} \approx 0.13$; (3) on the BCS side ($1/k_{F}a\approx -0.34$) in an isotropic trap, the oscillation frequency is $\omega_{B} \approx1.96\omega_{0}$ and the damping rate is $\Gamma_{B}/\omega_{B} \approx 0.006$. The above three examples indicate that, when the SO(2,1) symmetry is broken due to either strong anisotropy or away from resonant interaction, the oscillation frequency deviates from twice the trapping frequency and the damping rate increases.

\section{Breathing mode at large excitation amplitudes} \label{Sec:III}
\begin{figure} [htbp]
\includegraphics[width=8cm]{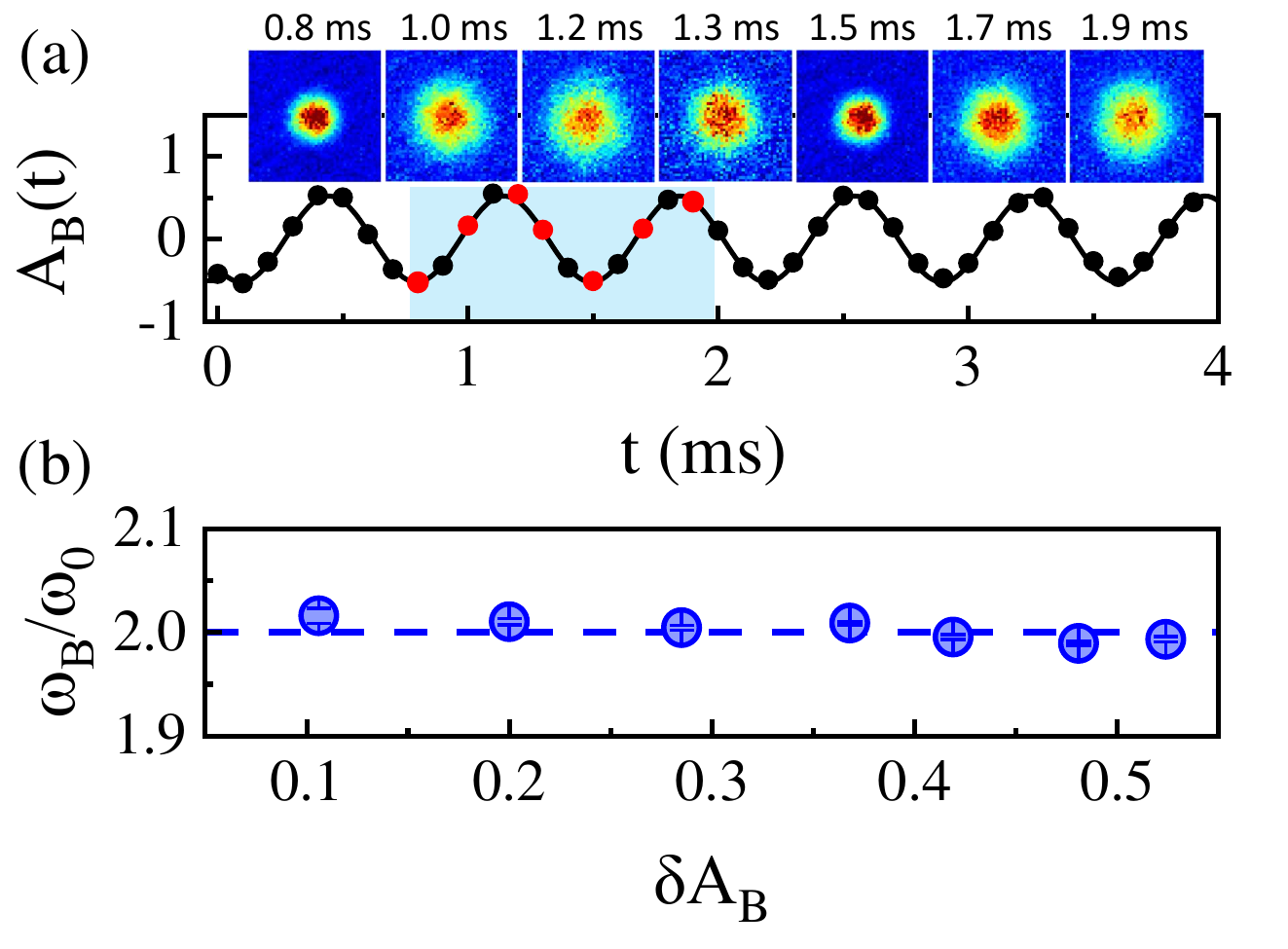}
\caption{Breathing mode at large excitation amplitudes. (a) The breathing oscillation at a large excitation amplitude $\delta A_{B}=0.52$. The error bar is the standard deviation of three measurements. The black solid curve is a sinusoidal fit of the data, which gives the oscillation frequency $\omega_{B} = 2\pi\times1425(2)\text{ Hz}\approx 2\omega_{0}$. The inset shows typical atomic images ($230  \times 230 \ \mu \textrm{m}^{2}$) from 0.8 to 2.0 ms, corresponding to the red dots. (b) Breathing-oscillation frequency $\omega_{B}/\omega_{0}$ for different excitation amplitudes. The trap anharmonicity has been corrected. The blue dashed line indicates the value $\omega_{B}/\omega_{0}=2$. The error bar comes from the sinusoidal fitting. Here, the temperature is $T/T_{F}=0.24(1)$, the atom number is $N=1\times10^{5}$, and the trapping frequency is $\omega_{0}=2\pi\times 720$ Hz.}
\label{fig:Fig2}
\end{figure}

As predicted by the SO(2,1) symmetry, another unusual behavior of the breathing-mode frequency is its independence on the excitation amplitude. In Fig. \ref{fig:Fig2}(a), we increase the excitation amplitude $\delta A_{B}$ to as large as $0.52$ of the initial cloud size and the oscillation frequency remains still at $\omega_{B}\approx 2\omega_{0}$.  We also observe a constant breathing-mode frequency at $\omega_{B}\approx 2\omega_{0}$ for different excitation amplitudes in Fig. \ref{fig:Fig2}(b). The large-amplitude oscillation corresponds to the coherent superposition of conformal-tower states (Fig. \ref{fig:Fig1}(b)) and goes beyond the linear excitation regime studied before~\cite{Castin1996, Stringari1996}. The SO(2,1) symmetry induces the decoupled breathing mode and consequently results in the independence of the oscillation frequency on the excitation amplitude. Away from resonant interaction with the SO(2,1) symmetry broken, amplitude-dependent frequency shift and mode coupling effect arise, which has been observed in an anisotropic trap~\cite{Pitaevskii1997PRAmodecoupling, Foot2000PRLmodecoupling, Foot2001PRLBeliaevcoupling}.
 %which has been explored around the equilibrium by linearizing the equations of motion \cite{Castin1996, Stringari1996} and exhibits the amplitude-dependent frequency shift and even mode coupling \cite{Pitaevskii1997PRAmodecoupling, Foot2000PRLmodecoupling, Foot2001PRLBeliaevcoupling}.

The insensitivity of the frequency to the amplitude suggests that this breathing motion is an interacting analogue of the Boltzmann breather. In fact,  this connection can be strengthened by the observation that the Wigner function for the unitary Fermi gas satisfies the streaming term of the Boltzmann equation \cite{Maki2024arXiv}, similar to how the Boltzmann breather was first predicted. We also note that similar physics has been observed in weakly interacting 2D quantum gases \cite{Dalibard2002PRL, Vogt2012} due to the classical SO(2,1) symmetry, while for the strongly interacting 2D Fermi gas, the damping rate $\Gamma_{B}/\omega_{B}$ increases by about ten times from its value in the weak interacting limit and the oscillation frequency shifts up away from twice the trapping frequency due to quantum anomaly \cite{Peppler2018, Holten2018}.
%from 0.0015 in the weakly interacting regime to 0.01 in the strongly interacting regime, and the oscillation frequency shifts up away from twice the trapping frequency \cite{Peppler2018, Holten2018}.

\section{Robustness of the SO(2,1) symmetry in 3D} \label{Sec:IV}
\begin{figure} [htbp]
\includegraphics[width=8.5cm]{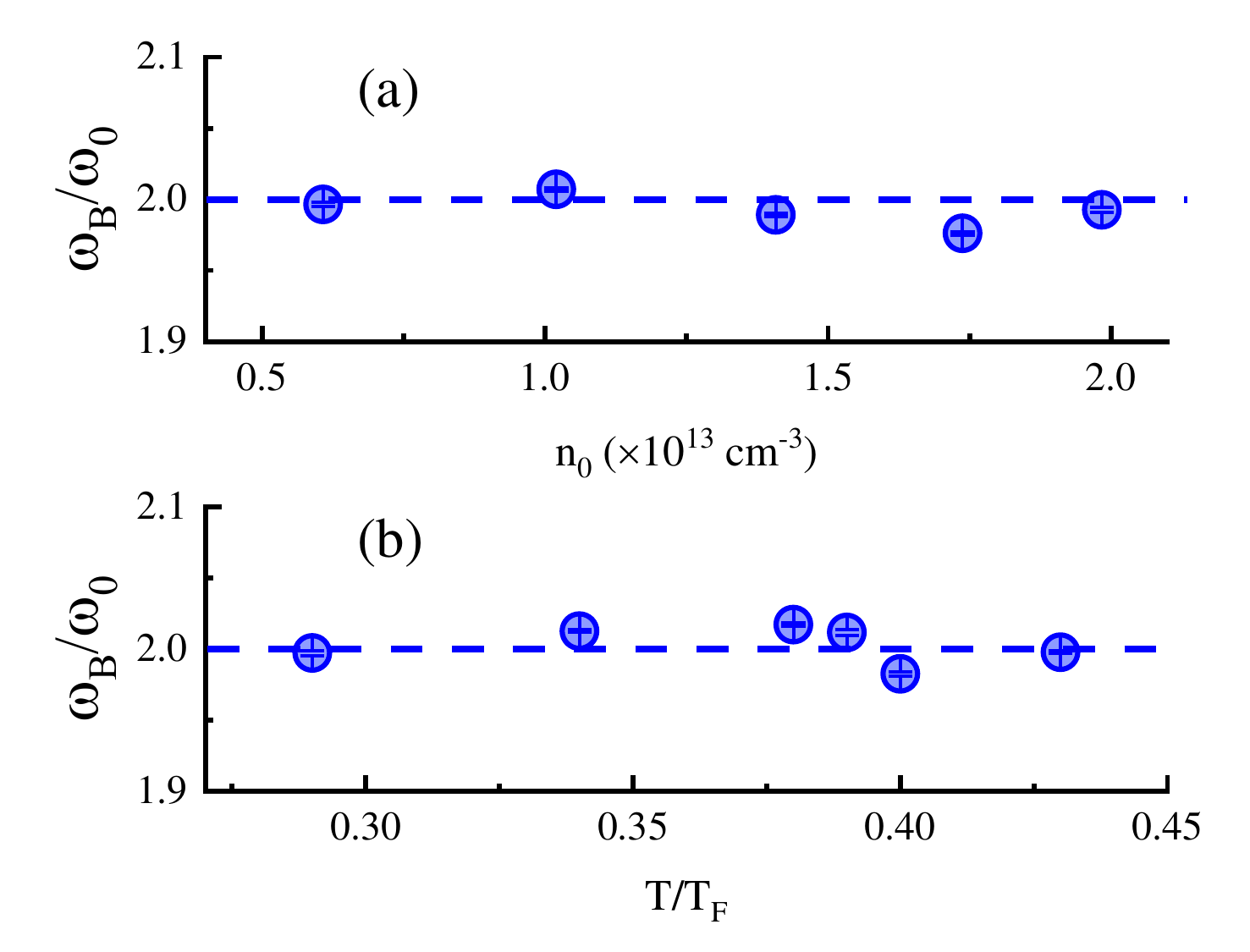}
\caption{Oscillation frequencies of the breathing mode at different atomic densities and temperatures. (a) The normalized breathing-mode frequency $\omega_{B}/\omega_{0}$ versus the central atomic density $n_{0}$.  Here $T/T_{F} = 0.29$. (b) $\omega_{B}/\omega_{0}$ versus the reduced temperature $T/T_{F}$. Here $n_{0} = 1.9 \times 10^{13} \ \textrm{cm}^{-3}$. $\omega_{0}$ is the trapping frequency and $T_{F}$ is the Fermi temperature. The error bar is the fitting uncertainty of the damped sinusoidal function. The blue dashed lines on both panels denote the value $\omega_{B}/\omega_{0} = 2$.}
\label{fig:Fig3}
\end{figure}

\begin{figure} [htbp]
\includegraphics[width=8.5cm]{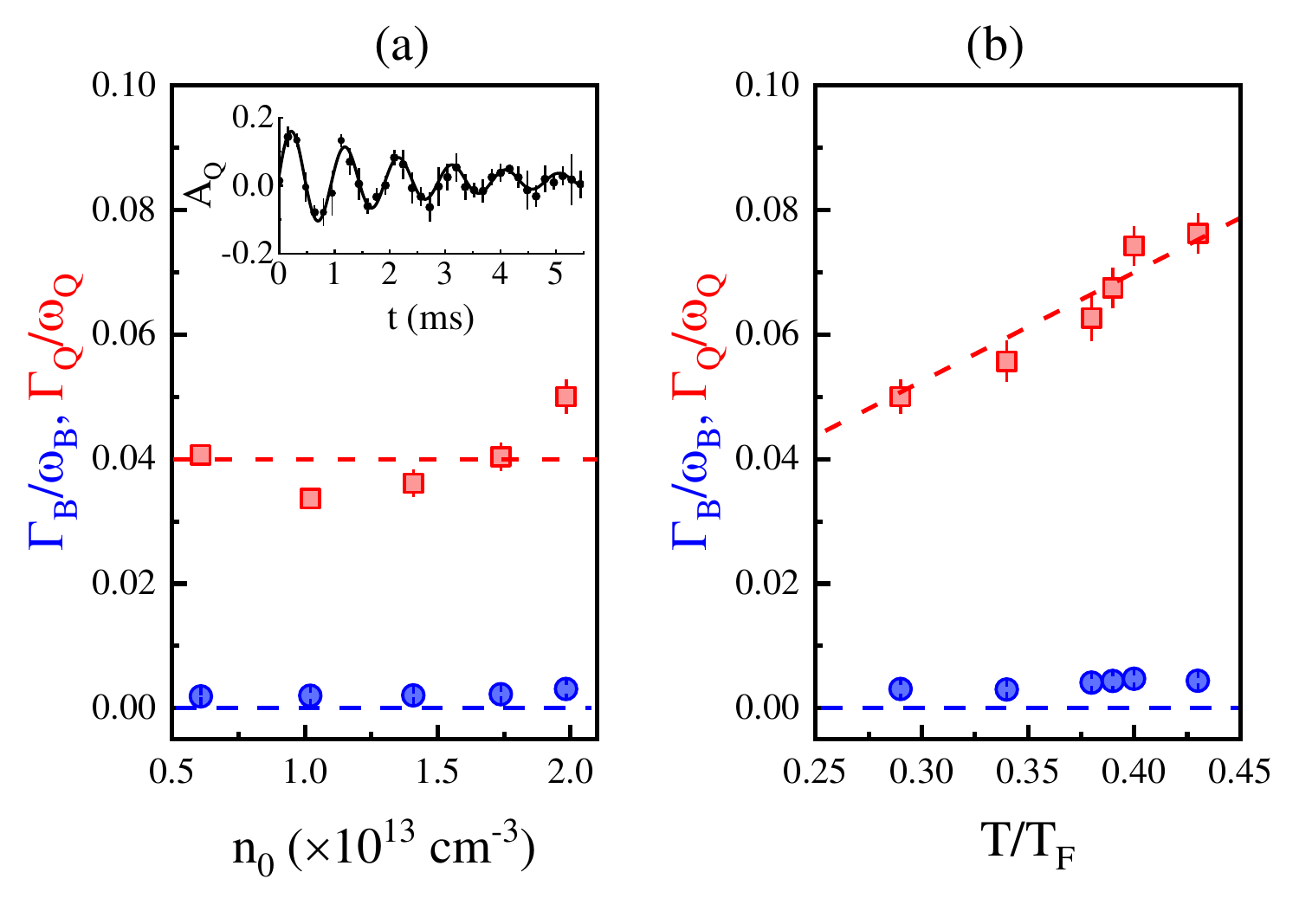}
\caption{Damping rates of breathing and quadrupole modes at different atomic densities and temperatures. (a) Breathing-mode damping rate $\Gamma_{B}/\omega_{B}$ versus the central atomic density $n_{0}$ (blue circles). The blue dashed line denotes the zero value. As a comparison, the quadrupole-mode damping rates $\Gamma_{Q}/\omega_{Q}$ are also shown (red squares).
%$\omega_{B} \approx 2 \omega_{0}$ and $\omega_{Q} \approx \sqrt{2} \omega_{0}$ are the oscillation frequencies of the breathing and quadrupole modes, respectively.
The dashed red line denotes the average value 0.04. The error bar is the fitting uncertainty of the damped sinusoidal function. Here $T/T_{F} = 0.29$. The inset shows a typical oscillation of the quadrupole mode. (b) $\Gamma_{B}/\omega_{B}$ versus the temperature $T/T_{F}$. The blue dashed line denotes the zero value and the red dashed line for linearly fitting the data of $\Gamma_{Q}/\omega_{Q}$. Here $n_{0} = 1.9 \times 10^{13} \ \textrm{cm}^{-3}$.}
\label{fig:Fig4}
\end{figure}

To demonstrate the robustness of the SO(2,1) symmetry, we study the breathing mode at different atomic densities and temperatures. When the central atomic density $n_{0}$ varies from $6\times10^{12}$ to $2\times10^{13} \ \textrm{cm}^{-3}$, or the temperature $T/T_{F}$ changes from  $0.29$ to $0.44$, the oscillation frequency of the breathing mode is always equal to twice the trapping frequency (see Fig. \ref{fig:Fig3}), and the damping rate remains always close to zero (see Fig. \ref{fig:Fig4}). This independence of atomic density and temperature is in direct contrast with the 2D situation, where the increase of the atom number or finite-temperature thermal fluctuation will induce the system to enter 3D regime, resulting in a large damping rate and downward shift of the oscillation frequency \cite{Peppler2018, Holten2018}. Previously, the temperature-dependent variation of the oscillation frequency and damping rate has been extensively studied in an anisotropic trap \cite{Ketterle1998PRLexcitation, Wieman1997PRLexcitation, Thomas2004PRLsuperfluidity, kinast2005damping, Grimm2007PRLcollective, riedl2008collective, Grimm2013PRLsuperfluid}. Here, the constant oscillation frequency $\omega_{B}/\omega_{0} \approx 2$ and nearly zero damping rate are protected by the SO(2,1) symmetry.
%To maintain a system in a 2D, the axial trapping potential energy should be much larger than the chemical potential, i.e., $\hbar \omega_{z}\gg \mu$.
%It has been shown  \cite{Peppler2018, Holten2018, Murthy2019} that
%toward to the 2D ground-state threshold value $N_{2D}$,
%the system will evolve from 2D to 3D with the increase of the atom number, and subsequently the breathing-mode frequency would shift down and the damping rate increases greatly. Also, the thermal fluctuation at finite temperature
%at about $T/T_{F}=0.2$
%in 2D will induce axial excitations, which has large effects on the frequency and damping rate of the breathing mode \cite{Peppler2018, Holten2018}.
%The 3D system does not suffer from the dimensional crossover, and consequently the damping rate and frequency of the breathing mode are robust against the fluctuations in atomic number and temperature.

\section{Factors affecting the residual damping} \label{Sec:V}
The damping rate of the breathing oscillation is very small, but not zero. Next we analyze the factors that affect the small residual damping of the breathing mode. We also excite the quadrupole mode $\left(\ell=2,\left|m\right|=2\right)$ by applying a sinusoidal asymmetric modulation of the trap in the $x-y$ plane at the frequency $\sqrt{2}\omega_{0}$ for four periods. The atomic cloud sizes oscillate out of phase along the $x$ and $y$ axes, while the width along the $z$ axis keeps nearly constant. The quadrupole oscillation is described by $A_{Q}(t)\equiv \frac{\left\langle r_{x}^{2}\right\rangle \left(t\right)-\left\langle r_{y}^{2}\right\rangle \left(t\right)}{\left\langle\left\langle r_{x}^{2}\right\rangle (t)+\left\langle r_{y}^{2}\right\rangle (t)\right\rangle}$.
%\begin{equation}
%A_{Q}(t)\equiv \frac{\left\langle r_{x}^{2}\right\rangle \left(t\right)-\left\langle r_{y}^{2}\right\rangle \left(t\right)}{\left\langle\left\langle r_{x}^{2}\right\rangle (t)+\left\langle r_{y}^{2}\right\rangle (t)\right\rangle}.
%\end{equation}
A typical dataset for $A_{Q}(t)$ is shown in the inset of Fig.~\ref{fig:Fig4}(a),
%Our measurements confirm that
where the oscillation frequency is $\omega_{Q}=2\pi\times1033(6)\text{ Hz}\approx\sqrt{2}\omega_{0}$ and
%Unlike the breathing mode, however,
the quadrupole mode decays rapidly after several milliseconds with a damping rate $\Gamma_{Q}=252(51)$ $\text{s}^{-1}$, in direct contrast to the breathing mode.

The observed damping rates of the breathing and quadrupole modes can be analyzed using the hydrodynamic theory \cite{Elliott2014}.
%We adopt a time-dependent scaling ansatz for the density profile $n\left({\bf r},t\right)=n_{0}\left(x/b_{x},y/b_{y},z/b_{z}\right)/(b_{x}b_{y}b_{z})$. It relates the density profile $n\left({\bf r},t\right)$ at time $t$ to that at equilibrium $n_{0}\left({\bf r}\right)$.
Starting from the Navier-Stokes equation, the scaling factors $b_{i}\left(t\right)$ obey the differential equations (for details, see Sec.~\ref{Sec:Appendix D}):
\begin{equation} \label{eq:HydrEq}
\frac{\ddot{b}_{i}}{b_{i}}=\frac{\omega_{i}^{2}}{b_{i}^{2}\left(\prod_{j}b_{j}\right)^{2/3}}-\omega_{i}^{2}-\frac{\bar{\alpha}_{s}\hbar}{m\left\langle r_{i}^{2}\right\rangle _{0}b_{i}^{2}}\sigma_{ii},
\end{equation}
where $i,j=x,y,z$, $\bar{\alpha}_{s}\equiv N^{-1}\int n_0(r)\alpha_{s}(n)dr$ is the trap-averaged shear viscosity coefficient, and $\sigma_{ii}=2\dot{b}_{i}/b_{i}-\left(2/3\right)\sum_{j}\dot{b}_{j}/b_{j}$. The bulk viscosity is set to zero as it vanishes for scale invariant interactions \cite{Son2007}. When the harmonic trap is isotropic ($\sigma_{ii} = 0$), the shear viscosity does not couple to the isotropic motion and then the breathing mode is undamped. In an anisotropic trap ($\sigma_{ii} \neq 0$), the shear viscosity $\bar{\alpha}_{s}$ will couple to the breathing mode, producing damping.
%For a small oscillation amplitude, Eq. (\ref{eq:HydrEq}) is linearized.
%by setting $b_{i}\approx 1+\epsilon_{i}$ with $\epsilon_{i} \ll 1$. Supposing the anisotropic trap has cylindrical symmetry, in particular we choose $\epsilon_{x} = \epsilon_{y} \neq \epsilon_{z}$ for the breathing mode, and $\epsilon_{x} = -\epsilon_{y}$ and $\epsilon_{z}=0$ for the quadrupole mode.
Expanding the solution for a small asphericity $\delta$ up to the second order gives the following estimates for the damping rates (see Sec.~\ref{Sec:Appendix D}):
\begin{equation} \label{eq:GammaQ}
\frac{\Gamma_{B}}{\omega_{B}}\approx\frac{16\gamma\delta^2}{9+36\gamma^2}, \ \quad \frac{\Gamma_{Q}}{\omega_{Q}}= \frac{\gamma}{\sqrt{2}} = \frac{\hbar\bar{\alpha}_{s}}{\sqrt{2}m\omega_0\left\langle r_{i}^{2}\right\rangle _{0}},
\end{equation}
\noindent where $\left\langle r_{i}^{2}\right\rangle_{0}$ ($i=x,y,z$) is the mean-square cloud radius along the $i$th axis without excitation in the trap. According to Eq. (\ref{eq:GammaQ}), $\bar{\alpha}_{s}$ can be extracted by measuring the damping of the quadrupole mode, which can be further used to calculate how the asphericity $\delta$ affects the damping of the breathing mode.

% At different atomic densities and temperatures, the damping rates of the breathing mode always remain close to zero, while those of the quadrupole mode are much larger and show a dependence on the atomic density and temperature.
In Fig. \ref{fig:Fig4}(a)
%the central atomic density $n_{0}$ varies from $6\times10^{12}$ to $2\times10^{13} \ \textrm{cm}^{-3}$
with a constant temperature $T/T_{F} =0.29$,
%, In this case
both $\bar{\alpha}_{s}$ and $\omega_{0}\left\langle r_{i}^{2}\right\rangle _{0}=\hbar(3N)^{1/3}/(3m) E/E_{F}$ remain unchanged. According to Eq. \eqref{eq:GammaQ}, $\Gamma_{Q}/\omega_{Q}$ should also be constant. The red dashed line in Fig. \ref{fig:Fig4}(a) denotes the average value of $\Gamma_{Q}/\omega_{Q}\approx 0.04$ which corresponds to $\bar{\alpha}_{s}=0.77$. In Fig. \ref{fig:Fig4}(b)
%the temperature $T/T_{F}$ changes from  $0.29$ to $0.44$
with a nearly constant central atomic density $n_{0}= 1.9\times10^{13} \ \textrm{cm}^{-3}$,
%, In this case
one expects $\bar\alpha_{s}\propto(T/T_{F})^{3}$ and $\langle r_{i}^{2}\rangle_{0}\propto(T/T_{F})^{2}$ \cite{cao2011universal}. Thus Eq.~\eqref{eq:GammaQ} states that $\Gamma_{Q}/\omega_{Q}$ should variy linearly with temperature $T/T_{F}$. A linear fit to the data in Fig.~\ref{fig:Fig4}(b) provides
%good agreement with
$\Gamma_{Q}/\omega_{Q}=0.175(4) T/T_{F}$.

\begin{figure} [htbp]
\includegraphics[width=8.5cm]{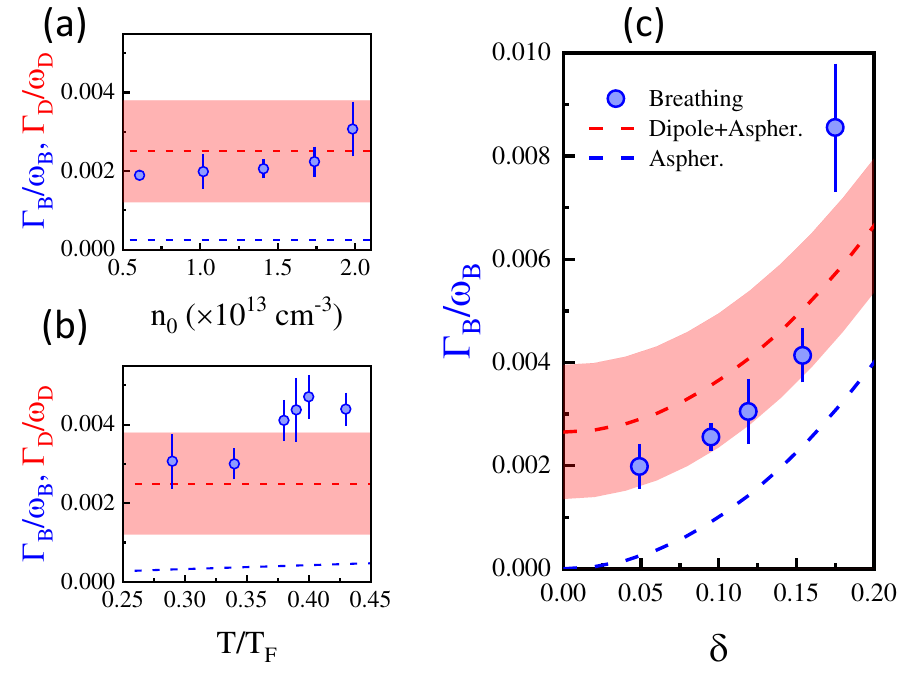}
\caption{Residual damping rate of the breathing mode. (a) and (b) are the enlarged views of the residual damping rates $\Gamma_{B}/\omega_{B}$ of the breathing mode in Fig. \ref{fig:Fig4}(a) and \ref{fig:Fig4}(b), respectively. Here the asphericity is $\delta=4.9\%$. The blue dashed lines denote the asphericity corrections of $\Gamma_{B}/\omega_{B}$ calculated with Eq. (\ref{eq:GammaQ}) and are in the order of $10^{-4}$.
%The anharmonicity correction is in the order of $10^{-4}$ and bulk viscosity correction is in the order of $10^{-6}$ (not shown here).
The damping rate $\Gamma_{D}/\omega_{D}$ of the dipole mode is 0.0025(13), where the average value is represented with the red dashed line and the uncertainty is indicated with the red shadow. In (c), the asphericity is increased up to about $18\%$, showing its obvious effect on the damping rate. The asphericity contribution is denoted with the blue dashed curve calculated with Eq. (\ref{eq:GammaQ}). The sum of the dipole-mode damping rate and asphericity contribution is represented by the red dashed curve with the red shadow. The error bar is the fitting uncertainty of the damped sinusoidal function.}
\label{fig:Fig5}
\end{figure}

Figure \ref{fig:Fig5} shows the residual damping rates of the breathing-mode in a very small range.
%Using $\bar{\alpha}_{s}$ obtained from the quadrupole-mode measurements,
%(red dashed lines in Fig. \ref{fig:Fig3}),
With Eq. (\ref{eq:GammaQ}) we can calculate the breathing-mode damping rate caused by the asphericity: for constant temperature at
different atomic densities, it is given by $\Gamma_{B}/\omega_{B} = 2.5\times 10^{-4}$ (Fig. \ref{fig:Fig5}(a)); for a fixed density at
different temperatures in the range $T/T_{F} = (0.29 - 0.44)$,
it increases from $3.1\times 10^{-4}$ to $4.6\times 10^{-4}$ (Fig. \ref{fig:Fig5}(b)).
%So the asphericity correction of the breathing-mode damping rate is about one order smaller than the experimental measurement.
%There could also be a contribution to the damping from the  anharmonicity of the trap.
We also estimate the effects of the trap anharmonicity and bulk viscosity. Since the trapping frequency is large with $\omega_{0} = 2\pi\times (501 \sim 1121) \ \textrm{Hz} $,
the maximal correction of the damping rate from the trap anharmonicity is about $1.5 \times 10^{-4}$ (see Sec.~\ref{Sec:Appendix D}).
%Another source of damping could be from finite interaction strengths
When the magnetic field is not exactly at the Feshbach resonance, the bulk viscosity is no longer zero. Assuming the off resonance to be 2 G,
the damping rate arising from bulk viscosity is
%no longer zero,
%and the damping rate caused by this factor is calculated to be
about $6 \times 10^{-6}$ (see Sec.~\ref{Sec:Appendix D}). Overall, the influences of the asphericity, anharmonicity and bulk viscosity are about one to three orders smaller than those observed in our experiment.
%too small to explain

The final source of the damping is the technical fluctuations, such as variations of the optical intensity, beam position, stray magnetic fields, and so on. We estimate these effects by measuring the dipole mode (see Fig. \ref{fig:sfig2}). The average value of the dipole-mode damping rate $\Gamma_{D}/\omega_{D}$ is 0.0025 and its uncertainty is 0.0013, which are also shown in Fig. \ref{fig:Fig5}(a) and (b). The residual damping rate of the breathing mode is on the same order of that of the dipole mode \cite{BreathingIncrease}.
So it is reasonable to claim that the breathing mode damping in our experiment is only limited by the weak fluctuations of the experimental parameters.
%The breathing-mode damping rate $\Gamma_{B}/\omega_{B}$ seems to increase with the atomic density and temperature. But we couldn't draw any concrete conclusions because the uncertainty in the dipole-mode damping rate is too large.

As a final check, in Fig. \ref{fig:Fig5}(c), we increase the asphericity $\delta$ to study its effect on the damping rate. The asphericity is controlled by adjusting the relative powers of two laser beams $P_{x}$ and $P_{y}$ (Fig. \ref{fig:Fig1}(a)). The atomic density and temperature are fixed at $n_{0}=6\times10^{12} \ \textrm{cm}^{-3}$ and $T/T_{F} =0.29$, respectively. We assume that the technical fluctuation contributes equally to dipole and breathing modes. A larger asphericity would lead to an increased damping rate of the breathing mode, which can be estimated by Eq. (\ref{eq:GammaQ}) after obtaining the shear viscosity $\bar{\alpha}_{s}$ from the quadrupole-mode measurement. The sum of the dipole-mode damping and asphercity correction provides a satisfactory agreement with the experimental results.

%The sum of the two damping rates (technical fluctuation as characterised by the dipole mode and asphericity) provides a satisfactory agreement with the experimental results.

%The contribution of the asphericity can be calculated with Eq. (\ref{eq:GammaQ}) after obtaining the shear viscosity $\bar{\alpha}_{s}$ from the quadrupole mode.
%The sum of the asphericity corrections and dipole-mode damping rate qualitatively agrees with the measured damping rate of the breathing mode.
%The joint contribution of asphericity and dipole mode provides a satisfactory agreement with the experimental results.

\section{Conclusion and outlook}
In conclusion, we have experimentally realized a long-lived breathing oscillation in an isotropic unitary Fermi gas. This motion is protected by the SO(2,1) symmetry and is the interacting analogue of the Boltzmann breather. The SO(2,1) symmetry in 3D is demonstrated to be robust in the strongly interacting regime,
%against quantum anomaly and dimensional crossover, 
different from the 2D situation. The factors that affect the residual damping rate of the breathing mode are also clarified.
%We further analyze the factors that affect the residual damping rate of the breathing mode and find that it is only limited by weak technical fluctuations of the experiment.

Our work opens a new way to study the non-equilibrium dynamics in a strongly interacting quantum system.
%In particular,
%The controllable sphericity provides a unique opportunity to study universal dynamics predicted by conformal symmetry \cite{Maki2019, Maki2020, Maki2022, Zhou2020PRLgeometrizing, Zhou2020PRLecho}. With suppressed contribution of the shear viscosity in an isotropic system, it also paves the way for a clean measurement of the bulk viscosity away from the unitary region \cite{Enss2019, Hofmann2020PRAviscosity, Nishida2019AnnalsVisocosity} or in the case of a narrow Feshbach resonance \cite{Zhang2022PRLviscosity}.
The long-lived isotropic breather in 3D provides a unique opportunity to study the quenched dynamics, thermalization and hydrodynamics related to the conformal symmetry \cite{Maki2019, Maki2020, Maki2022}. It also provides the platform to study geometrized quantum dynamics with the SU(1,1) symmetry (the same type as the SO(2,1) symmetry) \cite{Zhou2020PRLgeometrizing, Zhou2020PRLecho}. 

Generally, the shear viscosity is much bigger than the bulk viscosity. To measure the bulk viscosity, one should suppress the effect of the shear viscosity. In a spherical trap, the relative transverse velocity is zero for an isotropic motion and subsequently the contribution of the shear viscosity is negligible \cite{Jiang2024PRL}. By probing the breathing mode, we can measure the bulk viscosity along the BEC-BCS crossover \cite{Enss2019, Hofmann2020PRAviscosity, Nishida2019AnnalsVisocosity} or at a narrow Feshbach resonance \cite{Zhang2022PRLviscosity} in the future.

% resonance when conformal symmetry is broken 

\section*{ACKNOWLEDGMENTS}
This work is supported by the National Key R \&D Program under Grant No. 2022YFA1404102, the National Natural Science Foundation of China under Grants No. U23A2073, No. 12374250, and No. 12121004, Chinese Academy of Sciences under Grant No. YJKYYQ20170025, the Natural Science Foundation of Hubei Province under Grant No. 2021CFA027, and Innovation Program for Quantum Science and Technology under Grant No. 2023ZD0300401. S.Z. is supported by HK GRF Grants No. 17304820 and No. 17313122, CRF Grant No. C7012-21G, and a RGC Fellowship Award No. HKU RFS2223-7S03.

\setcounter{secnumdepth}{3} %%% Set the number depth as 3, that means the subsubsection is given a digital number. "3" to subsubsection, "2" to subsection
	
\setcounter{equation}{0}

\setcounter{section}{0}

\renewcommand{\thesection}{APPENDIX \Alph{section}}

\section{Experimental methods} \label{Sec:Appendix A}

\setcounter{equation}{0}
\renewcommand{\theequation}{A\arabic{equation}}

\textbf{Preparation of the unitary Fermi gases in an isotropic trap.} 
We first prepare an ultrcold $^{6}$Li atomic Fermi gas with a balanced spin mixture $|F=1/2,m_{F}=\pm1/2\rangle$ in an elongated optical dipole trap (ODT). Two elliptic 1064-nm optical beams, propagating along the $x$ and $y$ axes ($P_{x}$ and $P_{y}$), respectively, form the isotropic trap. The elliptic cross-sectional aspect ratio of the optical beam is $\sqrt{2}$, and the waist radius along the minor axis is $60\ \mu$m. A magnetic field with a gradient $B'_{z}=1.05$ G/cm is applied along the $z$-axis to simultaneously compensate the gravitational force of the two spin components. A pair of Feshbach coils produces a homogeneous magnetic field of 832 G to tune the interatomic interaction to the unitary limit. We transfer the unitary Fermi gas to the isotropic trap with an efficiency of more than ${90\%}$, by lowering the power of the elongated ODT while simultaneously increasing the power of the isotropic ODT over a period of 25 ms.

\begin{figure}[htbp]
\includegraphics[width=8cm]{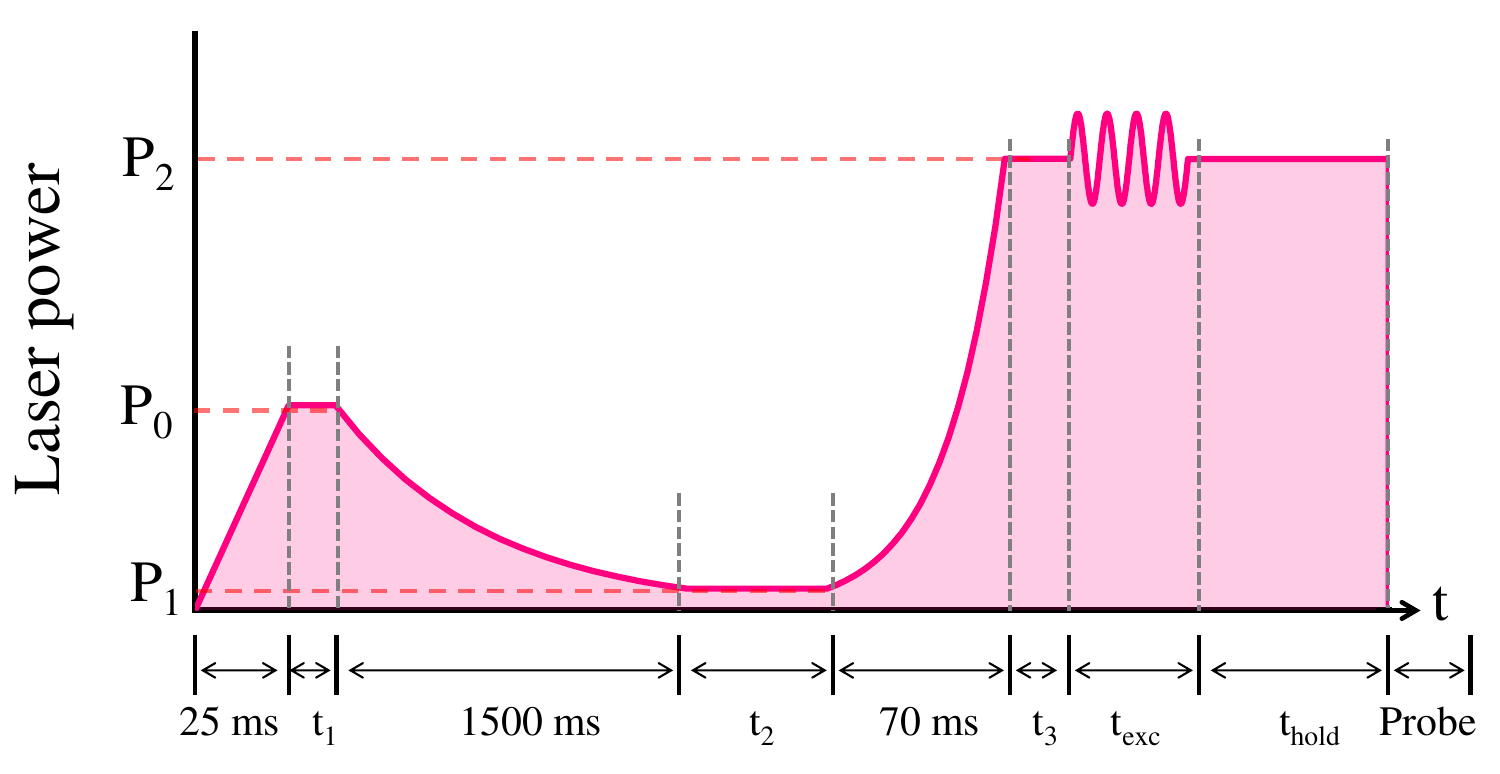}
\caption{Time sequence of the isotropic optical trap. The intial power of the isotropic optical trap is $P_{0}=380 \ \textrm{mW}$. Then the optical power is slowly decreased to $P_{1}$ during the evaporative cooling process and adiabatically increased to $P_{2}$ before exciting the collective mode. $t_{1, 2, 3}$ are the equilibrium times of the system: $t_{1}\approx20 \ \textrm{ms}$, $t_{2}\approx 100 \ \textrm{ms}$, $t_{3}\approx 30 \ \textrm{ms}$. The collective mode is excited by sinusoidally modulating the optical power at the frequency of the mode for four periods $t_{\textrm{exc}}$. After different holding times $t_{\textrm{hold}}$ in the trap, the atomic cloud is probed after a time of flight of 1 ms.}
\label{fig:Sfig1}
\end{figure}

\textbf{Time sequence of the isotropic optical trap.}
The time sequence of the isotropic optical trap is shown in Fig. \ref{fig:Sfig1}. The power of the isotropic optical trap is increased from zero to $P_{0}$ in 25 ms for initially loading the cold atoms ($P_{x}=P_{y}=P_{0}=380 \ \textrm{mW}$), slowly decreased to $P_{1}$ in 1500 ms for performing the evaporative cooling, and then adiabatically increased to $P_{2}$ in 70 ms for reducing the anharmonicity of the trap. $t_{1, 2, 3}$ are the equilibrium times of the system: $t_{1}\approx20 \ \textrm{ms}$, $t_{2}\approx 100 \ \textrm{ms}$, $t_{3}\approx 30 \ \textrm{ms}$. The temperature $T/T_{F}$ and central atom density $n_{0}$ are controlled by optical powers $P_{1}$ and $P_{2}$, respectively. The collective mode is excited by sinusoidally modulating the optical power at the frequency of the mode for four periods $t_{\textrm{exc}}$. After different holding times $t_{\textrm{hold}}$ in the trap, the atomic cloud is probed simultaneously along the $y$ and $z$ directions after a time of flight of 1 ms.

\textbf{Probing the widths of the atomic cloud.}
After different holding times $t$ in the trap, the atomic cloud is imaged with a time of flight of 1 ms, $t_{\textrm{TOF}}$=1 ms. Two laser beams with a frequency difference of 76 MHz, propagating along the $y$ and $z$ directions, respectively, are applied to detect the two spin states, obtaining the widths along three axes. We determine the mean-square cloud radius $\left\langle r_{i}^{2}\right\rangle \left(t\right)$ for $i=x,y,z$ by using a Gaussian fit of the density distribution, which is corrected by the finite-temperature Thomas-Fermi function at low temperature. The mean-square radius without excitation in the trap $\left\langle r_{i}^{2}\right\rangle_{0}$ can be determined by expanding the equilibrium gases and dividing the atomic radius with the expansion factor $\left(1+\omega_0^2t_{\textrm{TOF}}^2\right)$. For the breathing mode as in Fig. \ref{fig:Fig1}(c), %in the main text, 
there are five data sets each of which is composed of about 12 data points, and the average radius $\left\langle\sum_{i}\left\langle r_{i}^{2}\right\rangle (t)\right\rangle$ is obtained for each set. For the quadrupole mode as in the inset of Fig. \ref{fig:Fig4}(a), %of the main text, 
only one data set is shown due to its short lifetime and applied to get the average radius $\left\langle(\left\langle r_{x}^{2}\right\rangle (t)+\left\langle r_{y}^{2}\right\rangle (t))\right\rangle$.

\section{Measurement of the dipole mode} \label{Sec:Appendix B}

\setcounter{equation}{0}
\renewcommand{\theequation}{B\arabic{equation}}

\begin{figure*}[htbp]
\includegraphics[width=15cm]{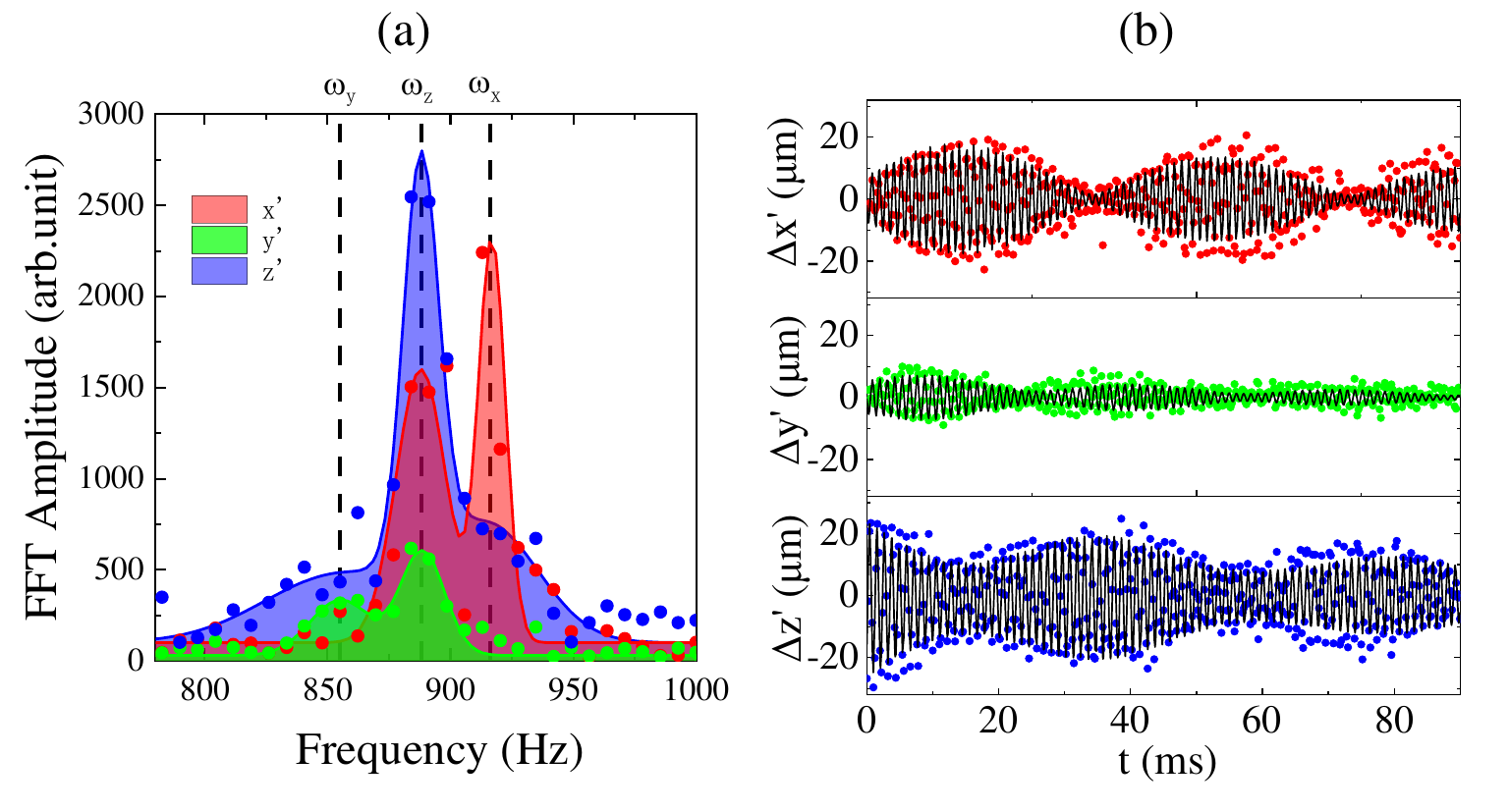}
\caption{Measurement of the dipole mode. (a) Fast-Fourier-transform (FFT) spectrum of the center-of-mass oscillations in the imaging coordinates ($x', y', z'$). The red, green, and blue curves with shadows represent the multiple-Gaussian fitting of the FFT data (colored circles) along the $x', y'$ and $z'$ directions, respectively. There is a small angle between the imaging coordinate system ($x', y', z'$) and the cold-atom coordinate system ($x, y, z$). The FFT spectrum along the $x'$ direction has two peaks of dipole-mode frequencies $\omega_{x}$ and $\omega_{z}$, $y'$ direction includes $\omega_{y}$ and $\omega_{z}$, and $z'$ direction includes $\omega_{x}$, $\omega_{y}$ and $\omega_{z}$, as shown with vertical dashed lines. (b) Center-of-mass oscillations versus the holing time $t$ in the trap. $\triangle x'$, $\triangle y'$ and $\triangle z'$ represent the center-of-mass displacements away from the equilibrium positions in the imaging coordinates ($x', y', z'$), where beat-frequency signals are obviously present. Each black solid curve denotes the fitting of a damped sinusoidal function including the FFT frequency components obtained in (a): along the $x'$ direction, the fitting function of Eq.~\eqref{eq:dipoleX} gives that $(A_{x}, A_{z})= [-11(1), 9(1)]$, $(\omega_{x}, \omega_{z})= 2\pi\times [917.1(5), 891.4(11)] \ \textrm{Hz}$, and $(\Gamma_{x}/\omega_{x}, \Gamma_{z}/\omega_{z})= [0.0015(4), 0.0028(12)]$; along the $y'$ direction, the fitting function of Eq.~\eqref{eq:dipoleY} gives that $(A_{y}, A_{z})= [-2.3(5), 5.7(5)]$, $(\omega_{y}, \omega_{z})= 2\pi\times [859.2(17), 889.1(20)] \ \textrm{Hz}$, and $(\Gamma_{y}/\omega_{y}, \Gamma_{z}/\omega_{z})= (0.0034(17), 0.0031(17))$; along the $z'$ direction, the fitting function of Eq.~\eqref{eq:dipoleZ} gives that $(A_{x}, A_{y}, A_{z})= [5(1), 4(1), 23(1)]$, $(\omega_{x}, \omega_{y}, \omega_{z})= 2\pi\times [913.5(8), 867.9(10), 889.9(12)] \ \textrm{Hz}$, and $(\Gamma_{x}/\omega_{x}, \Gamma_{y}/\omega_{y}, \Gamma_{z}/\omega_{z})= [0.0022(3), 0.0021(15), 0.0022(22)]$. The value in the parenthesis denotes the fitting uncertainty.}
\label{fig:sfig2}
\end{figure*}

A pulsed gradient-magnetic field is switched on to push the atomic cloud away from its equilibrium position. Then the center-of-mass oscillations versus the holing time $t$ in the trap are probed by two perpendicular CCDs. Because there is an angle between the imaging coordinate system ($x', y', z'$) and the cold-atom coordinate system ($x, y, z$), the dipole-mode oscillations in different directions of the cold-atom coordinate will be projected onto the center-of-mass oscillations in one direction of the imaging coordinate. The dipole-mode oscillations are shown in Fig. \ref{fig:sfig2}. A signal exhibiting the beat frequency emerges in the center-of-mass oscillation along each direction, which is also observed in Ref. \cite{altmeyer2007precision}. The fast-Fourier-transform (FFT) spectrum of the center-of-mass oscillations in the imaging coordinates ($x', y', z'$) clearly display the components of the dipole modes ($\omega_{x}$, $\omega_{y}$, $\omega_{z}$). We use a damped sinusoidal function composed of the FFT frequency components to fit the center-of-mass oscillation in the imaging coordinates ($x', y', z'$).

 The FFT spectrum along the $x'$ direction has two peaks of dipole-mode frequencies $\omega_{x}$ and $\omega_{z}$, $y'$ direction includes $\omega_{y}$ and $\omega_{z}$, and $z'$ direction includes $\omega_{x}$, $\omega_{y}$ and $\omega_{z}$. Then along the $x'$ direction, the fitting function is
\begin{equation} \label{eq:dipoleX}
\Delta x' (t)= \sum_{i} (A_{i}\sin(\omega_{i}t+\phi_{i})\exp(-\Gamma_{i}t)), \ \ (i=x, z),
\end{equation}

\noindent where $A_{i}$ is the oscillation amplitude, $\omega_{i}$ is the oscillation frequency, $\phi_{i}$ is the initial phase, and $\Gamma_{i}$ is the damping rate. We obtain $(A_{x}, A_{z})= [-11(1), 9(1)]$, $(\omega_{x}, \omega_{z})= 2\pi\times [917.1(5), 891.4(11)] \ \textrm{Hz}$, $(\phi_{x}, \phi_{z})= [-0.65(8), 5.01(14)]$ and $(\Gamma_{x}/\omega_{x}, \Gamma_{z}/\omega_{z})= [0.0015(4), 0.0028(12)]$. The values in the parentheses denote the fitting uncertainty.

Along the $y'$ direction, the fitting function is
\begin{equation} \label{eq:dipoleY}
\Delta y' (t)= \sum_{i} (A_{i}\sin(\omega_{i}t+\phi_{i})\exp(-\Gamma_{i}t)), \ \ (i=y, z).
\end{equation}

\noindent We find $(A_{y}, A_{z})= [-2.3(5), 5.7(5)]$, $(\omega_{y}, \omega_{z})= 2\pi\times [859.2(17), 889.1(20)] \ \textrm{Hz}$, $(\phi_{y}, \phi_{z})= [1.5(2), 2.8(1)]$ and $(\Gamma_{y}/\omega_{y}, \Gamma_{z}/\omega_{z})= [0.0034(17), 0.0031(17)]$.

Along the $z'$ direction, the fitting function is
\begin{equation} \label{eq:dipoleZ}
\Delta z' (t)= \sum_{i} (A_{i}\sin(\omega_{i}t+\phi_{i})\exp(-\Gamma_{i}t)), \ \ (i=x,y, z).
\end{equation}
\noindent We obtain $(A_{x}, A_{y}, A_{z})= [5(1), 4(1), 23(1)]$, $(\omega_{x}, \omega_{y}, \omega_{z})= 2\pi\times [913.5(8), 867.9(10), 889.9(12)] \ \textrm{Hz}$, $(\phi_{y}, \phi_{y}, \phi_{z})= [0.5(2), -1.9(4), -0.78(6)]$ and $(\Gamma_{x}/\omega_{x}, \Gamma_{y}/\omega_{y}, \Gamma_{z}/\omega_{z})= [0.0022(3), 0.0021(15), 0.0022(22)]$.

Averaging all the damping rates $\Gamma_{D}/\omega_{D}$ of the dipole mode obtained above, we get $\Gamma_{D}/\omega_{D} = 0.0025(13)$, which is used in Fig. \ref{fig:Fig5}. % of the main text.

\section{SO(2,1) Symmetry in a 3D unitary Fermi gas} \label{Sec:Appendix C}

\setcounter{equation}{0}
\renewcommand{\theequation}{C\arabic{equation}}

The unitary Fermi gas confined in an isotropic harmonic trap has the SO(2,1) symmetry, which has been introduced with details in Reference \cite{Werner2006}. A spherically trapped unitary Fermi gas of $N$ atoms is described by the Hamiltonian
\begin{align}
H=\sum_{j=1}^{N}\left(-\frac{\hbar^2}{2m}\nabla^2_j+\frac{1}{2}m\omega^2_0r^2_j\right),
\end{align}
where the interaction between atoms could alternatively be characterized by a short-range boundary condition obeyed by the wave function $\psi(r\rightarrow 0) \sim 1/r$. Let us introduce raising/lowing operators,
\begin{equation}
\begin{split}
L_+&=+\frac{3N}{2}+\hat{D}+\frac{H}{\hbar\omega_0}-\frac{m\omega_0}{\hbar}X^2,\\
L_-&=-\frac{3N}{2}-\hat{D}+\frac{H}{\hbar\omega_0}-\frac{m\omega_0}{\hbar}X^2,
\end{split}
\end{equation}
\noindent where $\hat{D}\equiv \vec{X}\cdot \partial _{\vec{X}}$ and $X^2=\sum_{j=1}^{N}r^2_j$. It is found that
\begin{equation}
\begin{split}
[H,L_+]&=2\hbar\omega_0 L_+,\\
[H,L_-]&=-2\hbar\omega_0 L_-,\\
[L_+,L_-]&=-4\frac{H}{\hbar\omega_0},
\end{split}
\end{equation}
which satisfies the algebra of SO(2,1). Repeated action $L_+$ on the lowest ground state $|\psi_0\rangle$ with energy $E_0$, we find
\begin{align}
H\underbrace{L_{+}...L_{+}}_{n}|\psi_0\rangle=(E_0+2n\hbar\omega_0)\underbrace{L_{+}...L_{+}}_{n}| \psi_0\rangle.
\end{align}
Also, repeated action $L_{-}$ on the state $|\psi\rangle$ with energy $E$, we find
\begin{align}
H\underbrace{L_{-}...L_{-}}_{n}|\psi\rangle=(E-2n\hbar\omega_0)\underbrace{L_{-}...L_{-}}_{n}|\psi\rangle.
\end{align}
Obviously an infinite number of excitations with energies $2n\omega_0$ exists, which we will identify with the breathing modes of the system.

%From the general theory of Lie algebras, one may form the so-called Casimir operator,
%\begin{align}
%\hat{\mathcal{C}}=H^2-\frac{1}{2}(\hbar\omega_0)^2(L_+L_-+L_-L_+),
%\end{align}
%which commutes with $H$ and $L_{\pm}$. So, $\hat{\mathcal{C}}$ is a scalar within each ladder. For the ground energy ladder, we have
%\begin{align}
%\hat{\mathcal{C}}|\psi_0\rangle=E_0\left(E_0-2\hbar\omega_0\right)|\psi_0\rangle.
%\end{align}

\section{Calculation on the Collective oscillations and damping} \label{Sec:Appendix D}

\setcounter{equation}{0}
\renewcommand{\theequation}{D\arabic{equation}}

We consider collective oscillations of a unitary Fermi gas in the framework of hydrodynamic theory. The basic equations of hydrodynamic theory are the continuity equation \citep{Elliott2014}
\begin{equation}
\frac{\partial n}{\partial t} + \nabla \cdot \left( n \mathbf{v} \right)\\ = 0,
\end{equation}
and Euler's equation
\begin{multline}
\left[ n \frac{\partial}{\partial t} + n \left( \mathbf{v} \cdot \nabla \right) \right] v_{i} = -\frac{1}{m} \frac{\partial p}{\partial r_{i}} - \frac{n}{m} \frac{\partial V_{\text{ext}}}{\partial r_{i}} \\
+ \frac{1}{m} \sum_{j} \frac{\partial}{\partial r_{j}} \left( \eta \sigma_{ij} + \zeta_B \sigma' \delta_{ij} \right),
\label{eq:2B21}
\end{multline}
where \( n\left( \mathbf{r}, t \right) \) is the atom density, \( \mathbf{v}\left( \mathbf{r}, t \right) \) is the velocity field, \( p\left( \mathbf{r}, t \right) \) is the local pressure, \( V_{\text{ext}}\left( \mathbf{r}, t \right) \) is the external potential, and \( \sigma_{ij} = \partial v_{i}/\partial r_{j} + \partial v_{j}/\partial r_{i} - 2 \delta_{ij} \nabla \cdot \mathbf{v}/3 \) is the shear stress tensor and \( \sigma' = \nabla \cdot \mathbf{v} \). Here, \( \eta \) and \( \zeta_B \) are the shear and bulk viscosities, respectively.

To solve the hydrodynamic equation, we adopt an ansatz for the density profile at time \( t \) \citep{Elliott2014},
\begin{equation}
n\left(x, y, z, t\right) = \frac{1}{b_{x}b_{y}b_{z}} n_{0}\left( \frac{x}{b_{x}}, \frac{y}{b_{y}}, \frac{z}{b_{z}} \right), \label{eq:2B26}
\end{equation}
which relates the density profile \( n\left( \mathbf{r}, t \right) \) at time \( t \) to the equilibrium density profile \( n_0\left( \mathbf{r} \right) \), with the time dependence entirely contained in the scaling factors \( b_{i}(t) \) for \( i = x, y, z \). Inserting this ansatz into the hydrodynamic equations, we find that the scaling factors satisfy 
\begin{equation}
\frac{\ddot{b}_{i}}{b_{i}} = \frac{\omega_{i}^{2}}{b_{i}^{2} \left( b_{x} b_{y} b_{z} \right)^{2/3}} - \omega_{i}^{2} - \frac{3\hbar \omega_{i}^{2}}{2 \mathcal{E}_{ho}^{(0)} b_{i}^{2}} \left( \bar{\alpha}_s \sigma_{ii} + \bar{\alpha}_B \sigma' \right),
\label{eq:2B31}
\end{equation}
where \( \omega_i \) is the trapping frequency along the \( i \)-th axis, \( \bar{\alpha}_{s,B} \equiv N^{-1} \int n_0 \alpha_{s,B} \, d\mathbf{r} \) is the trap-averaged shear (bulk) viscosity, and \( \mathcal{E}_{ho}^{(0)} \equiv \sum_{i} \left\langle m \omega_{i}^{2} r_{i}^{2}/2 \right\rangle_0 \) is the initial total potential energy. The bulk viscosity typically vanishes for unitary Fermi gases, so \( \bar{\alpha}_B = 0 \) \citep{Son2007}.

For small oscillations, Eq. (\ref{eq:2B31}) can be linearized by setting \( b_{i}(t) \approx 1 + \epsilon_{i}(t) \) with \( \epsilon_{i}(t) \ll 1 \), leading to the equation
\begin{equation}
\frac{d^{2} \epsilon_{i}}{dt^{2}} + \frac{3 \bar{\alpha}_{s} \hbar \omega_{i}^{2}}{\mathcal{E}_{ho}^{(0)}} \frac{d}{dt} \left( \epsilon_{i} - \epsilon \right) + 2 \omega_{i}^{2} \left( \epsilon_{i} + \epsilon \right) = 0, \label{eq:2B32}
\end{equation}
where \( \epsilon = \sum_{i} \epsilon_{i}/3 \). Introducing a geometric mean frequency \( \omega_0 = \left( \omega_{x} \omega_{y} \omega_{z} \right)^{1/3} \) and a dimensionless time \( \tau = \omega_0 t \), Eq. (\ref{eq:2B32}) becomes
\begin{equation} \label{eq:2B33}
\frac{d^{2} \epsilon_{i}}{d \tau^{2}} + 2 \gamma \left( \frac{\omega_{i}}{\omega_0} \right)^{2} \frac{d}{d \tau} \left( \epsilon_{i} - \epsilon \right) + 2 \left( \frac{\omega_{i}}{\omega_0} \right)^{2} \left( \epsilon_{i} + \epsilon \right) = 0, 
\end{equation}
for \( i = x, y, z \), where we have defined a dimensionless damping rate
\begin{equation}
\gamma \equiv \frac{3 \bar{\alpha}_s \hbar \omega_0}{2 \mathcal{E}_{ho}^{(0)}}. \label{eq:2B34}
\end{equation}
For a breathing mode in an isotropic trap with \( \omega_i = \omega_0 \), the cloud sizes along the three axes oscillate in phase, i.e., \( \epsilon_i = \epsilon \). In this case, the second term in Eq. (\ref{eq:2B33}) vanishes, and the system admits an undamped breathing solution with an oscillation frequency \( \omega_B = 2 \omega_0 \). The quadrupole mode studied in our experiment, on the other hand, exhibits oscillations with opposite phases along the \( x \) and \( y \) directions, i.e., \( \epsilon_x = -\epsilon_y \) and \( \epsilon_z = 0 \). From Eq. (\ref{eq:2B33}), this gives a solution for the quadrupole oscillation with a frequency \( \omega_Q = \sqrt{2} \omega_0 \) and a damping rate \( \Gamma_Q = \omega_0 \gamma \).

In our experiment, there is still a residual damping observed in the breathing oscillation. In the follows, we are going to discuss the possible reasons, such as trap asphericity, trap anharmonicity and bulk viscosity.

\subsubsection{Damping Due to Asphericity}

We assume cylindrical symmetry, where \( \omega_{x} = \omega_{y} \equiv \omega_{\perp} > \omega_{z} \). In the condition with $\epsilon_{x}=\epsilon_{y}\equiv \epsilon_{\perp}\neq\epsilon_{z}$, Eq.~(\ref{eq:2B33}) leads to two coupled equations. In the presence of slight asphericity, i.e., \( \delta = \left(\omega_{\perp} - \omega_{z}\right)/\omega_{0} \ll 1 \), expanding these equations up to the second order in \( \delta \), we obtain:
\begin{multline} \label{eq:M1}
\frac{d^{2}\epsilon_{\perp}}{d\tau^{2}} + 2\left(1 + \frac{2\delta}{3} + \frac{\delta^{2}}{3}\right) \gamma \frac{d}{d\tau}\left(\epsilon_{\perp} - \epsilon\right) \\
+ 2\left(1 + \frac{\delta}{3} + \frac{\delta^{2}}{9}\right)\left(\epsilon_{\perp} + \epsilon\right) = 0,
\end{multline}
\begin{multline} \label{eq:Q2}
\frac{d^{2}\epsilon_{z}}{d\tau^{2}} + 2\left(1 - \frac{4\delta}{3} + \frac{2\delta^{2}}{3}\right) \gamma \frac{d}{d\tau}\left(\epsilon_{z} - \epsilon\right) \\
+ 2\left(1 - \frac{2\delta}{3} + \frac{\delta^{2}}{9}\right)\left(\epsilon_{z} + \epsilon\right) = 0.
\end{multline}

By solving Eqs.~(\ref{eq:M1}) and (\ref{eq:Q2}) perturbatively up to the second order in \( \delta \), we obtain the damping rates for the two eigenmodes:
\begin{eqnarray}
\frac{\Gamma_{B}}{\omega_{0}} & \approx & \frac{32 \gamma \delta^2}{9 + 36 \gamma^2}, \label{eq:DampB} \\
\frac{\Gamma_{Q}}{\omega_{0}} & \approx & \gamma - \frac{2}{3} \gamma \delta + \frac{\gamma \left(-27 + 20 \gamma^2 \right)}{9 + 36 \gamma^2} \delta^2. \label{eq:DampQ}
\end{eqnarray}
The influence of asphericity on the damping of the quadrupole mode is negligibly small, so we have \( \Gamma_{Q}/\omega_{0} \approx \gamma \). It is important to note that the quadrupole mode described by Eqs.~(\ref{eq:M1}) and (\ref{eq:Q2}) corresponds to \( \left(l = 2, |m| = 0\right) \). The quadrupole oscillation excited in our experiment corresponds to \( \left(l = 2, |m| = 2\right) \) with \( \epsilon_{x} = -\epsilon_{y} \) and \( \epsilon_{z} = 0 \). Using the same method, we could calculate that the damping rate of the quadrupole oscillation measured in the experiment is nearly identical to that of the quadrupole mode corresponding to \( \left(l = 2, |m| = 0\right) \), as given by Eq.~(\ref{eq:2B33}). From the measured $\Gamma_{Q}/\omega_{Q}$, we get the value of $\gamma$. Then using Eq. (\ref{eq:DampB}) with $\delta=4.9\%$, we can calculate damping rate of the breathing mode, $\Gamma_{B}/\omega_{B}=2.5\times10^{-4}$, as shown in Fig. \ref{fig:Fig5}(a).

The temperature dependence of the damping rate for the breathing mode can be estimated experimentally as follows. From Eqs.~(\ref{eq:2B34}) and (\ref{eq:DampQ}), the damping rate of the quadrupole mode is given by:  
\begin{equation}
\frac{\Gamma_{Q}}{\omega_{Q}} \approx \frac{3^{4/3}m\omega_0 \bar{\alpha}_{s}}{2\sqrt{2}\pi\hbar n_0^{2/3}} \left(\frac{E_F}{E}\right)^2,
\end{equation}
where \( \omega_{Q} \) is the frequency of the quadrupole oscillation. Since \( E/E_F \propto T/T_F \) and \( \bar{\alpha}_{s} \propto \left(E/E_F\right)^{3/2} \left(T/T_F\right)^{3/2} \), we deduce that \( \Gamma_{Q}/\omega_{Q} \propto T/T_F \). In the main text, the data for \( \Gamma_{Q}/\omega_{Q} \) is fitted with the linear relation \( \Gamma_{Q}/\omega_{Q} = \xi T/T_F \), yielding \( \xi = 0.175(4) \). Using this result in Eq.~(\ref{eq:DampB}), we find that \( \Gamma_{B}/\omega_{B} \) increases from \( 3.1 \times 10^{-4} \) to \( 4.6 \times 10^{-4} \), as shown in Fig.~\ref{fig:Fig5}(b). Then the damping rate for the breathing mode induced by asphericity is found to be approximately one order of magnitude smaller than the experimentally measured values.

\subsubsection{Damping due to anharmonicity}

As the cloud size increases during excitation, anharmonicity becomes significant. This anharmonicity can contribute to the decay of collective oscillations and influence the oscillation frequencies. In our experiment, the trapping potential experienced by the atoms is described by  
\begin{equation} \label{eq:trap}
V_{\text{ext}}\left(\bf{r}\right) = U_{0}\left[2 - \left(e^{-2x^{2}/w_{\perp}^{2}} + e^{-2y^{2}/w_{\perp}^{2}}\right)e^{-2z^{2}/w_{z}^2}\right],
\end{equation}
where \( w_{\perp,z} \) are the beam waists, and \( U_{0} \) is the potential depth at the trap center. Expanding \( V_{\text{ext}}(\bf{r}) \) to fourth-order terms in \(\bf{r}\), we obtain  
\begin{multline}
\frac{V_{\text{ext}}\left({\bf r}\right)}{\hbar\omega_{0}} \approx \frac{1}{2}\left(\frac{r}{a_{\text{ho}}}\right)^{2} \\
- \frac{\beta}{2}\left[\left(\frac{x}{a_{\text{ho}}}\right)^{4} + \left(\frac{y}{a_{\text{ho}}}\right)^{4}\right] - \frac{\beta}{4}\left(\frac{z}{a_{\text{ho}}}\right)^{4} \\
- \frac{\beta}{2}\left(\frac{z}{a_{\text{ho}}}\right)^{2}\left(\frac{x}{a_{\text{ho}}}\right)^{2} - \frac{\beta}{2}\left(\frac{y}{a_{\text{ho}}}\right)^{2}\left(\frac{z}{a_{\text{ho}}}\right)^{2},
\end{multline}
where \( \beta = \hbar\omega_{0}/4U_{0} \) is the anharmonicity coefficient, \( a_{\text{ho}} = \sqrt{\hbar/m\omega_{0}} \) is the harmonic oscillator length, and the trap frequency is \( \omega_{0} = \sqrt{4U_{0}/mw_{\perp}^2} \). In our setup, the beam waist along the \( z \)-axis is \( \sqrt{2} \) times that in the transverse direction (\( w_{z} = \sqrt{2}w_{\perp} \)), resulting in a spherical trap. Using this anharmonic potential, Eq.~(\ref{eq:2B33}) leads to  
\begin{multline} \label{eq:anharmonic1}
\frac{d^{2}\epsilon_{\perp}}{d\tau^{2}} + 2\gamma\frac{d}{d\tau}\left(\epsilon_{\perp} - \epsilon\right) + 2\left(\epsilon_{\perp} + \epsilon\right) \\
- \frac{4\beta}{a_{\text{ho}}^2}\frac{\left\langle x^{4}\right\rangle _{0}}{\left\langle x^{2}\right\rangle _{0}}\left(2\epsilon_{\perp} + \epsilon\right) - \frac{2\beta}{a_{\text{ho}}^{2}}\frac{\left\langle x^{2}z^{2}\right\rangle _{0}}{\left\langle x^{2}\right\rangle _{0}}\left(4\epsilon - \epsilon_{\perp}\right) = 0,
\end{multline}
\begin{multline} \label{eq:anharmonic2}
\frac{d^{2}\epsilon_{z}}{d\tau^{2}} + 2\gamma\frac{d}{d\tau}\left(\epsilon_{z} - \epsilon\right) + 2\left(\epsilon_{z} + \epsilon\right) \\
- \frac{2\beta}{a_{\text{ho}}^{2}}\frac{\left\langle z^{4}\right\rangle _{0}}{\left\langle z^{2}\right\rangle _{0}}\left(2\epsilon_{z} + \epsilon\right) - \frac{4\beta}{a_{\text{ho}}^{2}}\frac{\left\langle x^{2}z^{2}\right\rangle _{0}}{\left\langle z^{2}\right\rangle _{0}}\left(4\epsilon - \epsilon_{\perp}\right) = 0. 
\end{multline}
In the absence of anharmonicity (\( \beta \rightarrow 0 \)), these equations reduce to those for a spherical trap. The equilibrium values \( \left<x^{2}\right>_{0}, \left<z^{2}\right>_{0}, \left<x^{4}\right>_{0}, \left<z^{4}\right>_{0}, \left<x^{2}z^{2}\right>_{0} \) are estimated using the 1D density profile of a unitary Fermi gas \cite{kinast2005damping}:  
\begin{equation}
n\left(x,T\right) \propto \text{Li}_{5/2}\left[-\exp\left(\frac{\mu/E_{F} - x^{2}/R_{x}^{2}}{T/\sqrt{\xi_{B}}T_{F}} \right)\right],
\end{equation}
where \( R_{x} = \xi_{B}^{1/4}(24N)^{1/6}a_{\text{ho}} \) is the Thomas-Fermi radius, \( \mu \) is the chemical potential, and \( \xi_{B} \approx 0.37 \) is the Bertsch parameter. Solving Eqs.~(\ref{eq:anharmonic1}) and (\ref{eq:anharmonic2}), we find the breathing mode damping rate due to anharmonicity:  
\begin{equation}
\frac{\Gamma_{B}}{\omega_{0}} = \frac{48\gamma}{9 + 36\gamma^{2}} \chi^{2}\beta^{2},
\end{equation}
where  
\begin{equation}
\chi = (3N)^{1/3} \frac{\text{Li}_{5}\left(-e^{\sqrt{\xi_{B}}\mu/k_{B}T}\right)}{\text{Li}_{4}\left(-e^{\sqrt{\xi_{B}}\mu/k_{B}T}\right)} \frac{T}{T_{F}}.
\end{equation}
It is indicated that the effect of anharmonicity decrease as the trap depth increases, and increases as the atomic cloud size increases. When the trapping frequency is $\omega_0=2\pi \times 501$ Hz, the anharmonicity correction of $\Gamma_B/\omega_B$ is about $1.5\times10^{-4}$. When the trapping frequency increases to $\omega_0=2\pi \times 1121$ Hz, the anharmonicity correction is only $6.7\times10^{-5}$. This means that, In our experiment, the anharmonicity correction of the breathing-mode damping rate is about one to two order smaller than the experimental measurement. 

\subsubsection{Damping due to bulk viscosity}

Bulk viscosity introduces additional damping to the breathing oscillations when the magnetic field is slightly detuned from resonance, which is likely the case in the experiment. In the high-temperature limit for a normal gas, bulk viscosity can be estimated using the virial expansion \cite{Enss2019PRLbulkviscosity}:
\begin{align}
\zeta_B(\mathbf{r}) \approx \frac{2\sqrt{2}\hbar}{9\pi} \lambda^{-3} v^2 \left[-1 - (1 + v^2) e^{v^2} \, \text{Ei}(-v^2)\right] z^2(\mathbf{r}),
\end{align}
where \( z(\mathbf{r}) = e^{\mu(\mathbf{r}) / (k_B T)} \) is the fugacity, with local chemical potential \( \mu(\mathbf{r}) \), \( \lambda = \sqrt{2 \pi \hbar^2 / (m k_B T)} \) is the thermal wavelength, and \( v = \lambda / (\sqrt{2\pi} a_0) \) is the dimensionless interaction strength, with the scattering length \( a_0 \). Here, \( \text{Ei}(\cdot) \) denotes the exponential integral.

The trap-averaged bulk viscosity is then given by
\begin{align}
\bar{\alpha}_B \approx \frac{2\sqrt{2}}{9\pi N} \lambda^{-3} v^2 \left[-1 - (1 + v^2) e^{v^2} \, \text{Ei}(-v^2)\right] \int z^2(\mathbf{r}) \, d\mathbf{r}.
\end{align}
The integral in \( \bar{\alpha}_B \) can be evaluated using the local density approximation (LDA), where \( \mu(\mathbf{r}) = \mu_0 - V_{\text{ho}}(\mathbf{r}) \) and \( \mu_0 \) is the chemical potential at the trap center. We then have
\begin{align}
\int z^2(\mathbf{r}) \, d\mathbf{r} = z_0^2 \int e^{-m \omega_0^2 r^2 / (k_B T)} \, d\mathbf{r} = (\sqrt{2} \pi)^3 \frac{a_{\text{ho}}^6}{\lambda^3} z_0^2,
\end{align}
which yields
\begin{align}
\bar{\alpha}_B \approx \frac{8 \pi^2}{9N} v^2 \left[-1 - (1 + v^2) e^{v^2} \, \text{Ei}(-v^2)\right] \left(\frac{a_{\text{ho}}}{\lambda}\right)^6 z_0^2.
\end{align}
Here, \( z_0 \) is determined by the total atom number \( N = \int n(\mathbf{r}) \, d\mathbf{r} \). The density distribution can also be obtained using the virial expansion \cite{Ho2004}:
\begin{align}
n(\mathbf{r}) \lambda^3 = 2 z(\mathbf{r}) \left[ 1 + 2^{3/2} b_2 z(\mathbf{r}) \right] - \frac{z^2(\mathbf{r})}{\sqrt{2}},
\end{align}
with the second virial coefficient \( b_2 = 1/2 \) in the unitary limit \cite{Ho2004PRLexpansion}. Thus, we have
\begin{align}
n(\mathbf{r}) \lambda^3 = 2 z(\mathbf{r}) + \frac{3 \sqrt{2}}{2} z^2(\mathbf{r}).
\end{align}
Integrating this equation over all space, we get
\begin{align}
N = 16 \pi^3 \left(\frac{a_{\text{ho}}}{\lambda}\right)^6 z_0 + 6 \pi^3 \left(\frac{a_{\text{ho}}}{\lambda}\right)^6 z_0^2.
\end{align}
This equation can be solved perturbatively in the high-temperature limit (\( \lambda \rightarrow 0 \)), yielding
\begin{align}
z_0 = \frac{N}{16 \pi^3} \left(\frac{\lambda}{a_{\text{ho}}}\right)^6 - \frac{3N^2}{2048 \pi^6} \left(\frac{\lambda}{a_{\text{ho}}}\right)^{12} + O\left(\frac{\lambda}{a_{\text{ho}}}\right)^{18}.
\end{align}
Finally, the trap-averaged bulk viscosity becomes
\begin{align}
\bar{\alpha}_B \approx \frac{N}{288 \pi^4} v^2 \left[-1 - (1 + v^2) e^{v^2} \, \text{Ei}(-v^2)\right] \left(\frac{\lambda}{a_{\text{ho}}}\right)^6.
\end{align}

In our experiment, assuming a magnetic field detuning of 2 G from the resonant interaction, we obtain \( \bar{\alpha}_B \approx 1.08 \times 10^{-4} \). Using Eq.~(\ref{eq:2B31}), the damping rate of the breathing mode caused by bulk viscosity is
\begin{align}
\frac{\Gamma_B}{\omega_B} \approx \frac{3 \hbar \bar{\alpha}_B}{4 m \omega_0 \langle r_x^2 \rangle_0}, \label{eq:2E91}
\end{align}
which gives approximately \( 6 \times 10^{-6} \) for conditions where \( T / T_F = 0.29 \) and \( n_0 \approx 1.9 \times 10^{13} \ \text{cm}^{-3} \). This suggests that the bulk viscosity correction to the breathing-mode damping rate is two to three orders of magnitude smaller than the experimental measurement.

%\bibliography{CollectiveRef}

\begin{thebibliography}{54}
\bibitem [{Rad()}]{RadiusOscillation}%
  \BibitemOpen
  \bibfield  {title} {\bibinfo {title} {The mean-square radius of the gas
  satisfies the following equation of motion: $\frac{d^2}{dt^2}\langle r^2
  \rangle(t) +4 \omega_{0}^2 \langle r^2 \rangle(t) = \frac{4E}{m}$, where
  {$E$} is the energy per atom and $m$ is the atomic mass},\ }\href@noop {} {\
  }\BibitemShut {NoStop}%
\bibitem [{Bre()}]{BreathingIncrease}%
  \BibitemOpen
  \bibfield  {title} {\bibinfo {title} {The breathing-mode damping rate
  $\gamma_{B}/\omega_{B}$ seems to increase with the atomic density and
  temperature. but we couldn't draw any concrete conclusions because the
  uncertainty in the dipole-mode damping rate is too large},\ }\href@noop {} {\
  }\BibitemShut {NoStop}%
\end{thebibliography}
%\bibliographystyle{unsrt}

\providecommand{\noopsort}[1]{}\providecommand{\singleletter}[1]{#1}%

\end{document}